\documentclass[journal]{jpp}

       \newcommand{\Ec}{ {\mathcal{E}} }

       \newcommand{\Hc}{ {\mathcal{H}} }
       
       \newcommand{\Jc}{ {\mathcal{J}} }
       
       \newcommand{\Lc}{ {\mathcal{L}} }

       \newcommand{\Oc}{ {\mathcal{O}} }

       \newcommand{\Sc}{ {\mathcal{S}} }

       \newcommand{\bfB}{{\boldsymbol B}}

       \newcommand{\bfE}{{\boldsymbol E}}

       \newcommand{\bfJ}{{\boldsymbol J}}

  \newcommand{\nn}{\nonumber}

  \newcommand{\nb}{\nabla}

  \newcommand{\bsv}{\boldsymbol{v}}

\usepackage{graphicx}
\usepackage{epstopdf, epsfig}
\usepackage[colorlinks]{hyperref}
\usepackage{amsfonts}
\usepackage{amssymb}
\usepackage[T1]{fontenc}
\usepackage{mathtools}
\usepackage{amsfonts}
\usepackage{dcolumn}
\usepackage{bm}
\usepackage[export]{adjustbox}
\usepackage{wrapfig}
\usepackage{lipsum}
\allowdisplaybreaks
\usepackage[titletoc,toc,title]{appendix}

\shorttitle{Chaotic and integrable magnetic fields in hybrid equilibria}
\shortauthor{D. A. Kaltsas, P. J. Morrison, G. N. Throumoulopoulos}

\title{Chaotic and integrable magnetic fields in one-dimensional hybrid Vlasov-Maxwell equilibria}

\author{Dimitrios A. Kaltsas\aff{1,2}
  \corresp{\email{d.kaltsas@uoi.gr}},
  Philip J. Morrison \aff{3}\\ \and
  George N. Throumoulopoulos \aff{1}}

\affiliation{\aff{1}Department of Physics, University of Ioannina, Ioannina GR 451 10, Greece.
\aff{2} Department of Physics, International Hellenic University, Kavala GR 654 04, Greece.
\aff{3}Department of Physics and Institute for Fusion Studies, The University of Texas at Austin, Austin, TX 78712, USA}
\date{2023}
\begin{document}
\maketitle

\begin{abstract}
The construction of kinetic equilibrium states is important for studying stability and wave propagation in collisionless plasmas.  Thus,  many studies over the past decades have been focused on calculating Vlasov-Maxwell equilibria using analytical and numerical methods. However, the problem of kinetic equilibrium of hybrid models is less studied, and self-consistent treatments often adopt restrictive assumptions ruling out cases with irregular and chaotic behavior,  although such behavior  is observed in spacecraft observations of space plasmas. In this paper, we develop a one-dimensional (1-D), quasineutral, hybrid Vlasov-Maxwell equilibrium model with kinetic ions and massless fluid electrons and derive associated solutions. The model allows for an electrostatic potential that is expressed in terms of the vector potential components through the quasineutrality condition. The equilibrium states are calculated upon solving an inhomogeneous Beltrami equation that determines the magnetic field, where the inhomogeneous term  is the current density of the kinetic ions and the homogeneous term represents the electron current density. We show that the corresponding 1-D system is Hamiltonian, with position playing the role of time,  and its trajectories have a regular, periodic behavior for ion distribution functions that are symmetric in the two conserved particle canonical momenta. For asymmetric distribution functions, the system is nonintegrable, resulting in irregular and chaotic behavior of the fields. The electron current density can modify the magnetic field phase space structure, inducing orbit trapping and the organization of orbits into large islands of stability. Thus the electron contribution can be responsible for the emergence of localized electric field structures that induce ion trapping. We also provide a paradigm for the analytical construction of hybrid equilibria using a rotating two-dimensional harmonic oscillator Hamiltonian, enabling the calculation of analytic magnetic fields and the construction of the corresponding distribution functions in terms of Hermite polynomials.
\end{abstract}


\section{Introduction}
%
In the study of collisionless plasmas, when dealing with scales comparable to ion characteristic time and length scales, it is essential to take into account ion kinetic effects. Of course, a fully kinetic approach that utilizes solutions of the Vlasov equation for both the ion and the electron component of the plasma provides a fundamental description. However, it is often the case that the plasma consists of suprathermal ions and thermal electrons, e.g., due to selective ion energization, or the electron scales are irrelevant, for example in the propagation of slow ion-acoustic waves. In such cases, a viable alternative is to use a hybrid model that treats only the ion component kinetically. A particularly successful model which has been utilized in order to study various processes that involve ion scales, is the hybrid Vlasov-Maxwell model with kinetic ions and massless fluid electrons \cite{Valentini2007,Tronci2010,Servidio2012,Cerri2016,Franci2017}. As this model has been used in several kinetic simulations to study physical processes, such as magnetic reconnection, kinetic turbulence and shearing flows, the problem of constructing consistent equilibrium states that can serve as reliable initial conditions in stability and dynamics calculations, arises. It is common practice in the literature to use Maxwellian and shifted Maxwellian distribution functions as initial conditions. However, in collisionless plasmas it is anticipated that the distribution functions of the kinetic species will deviate from the Maxwellian distribution. Using incorrect or off-equilibrium initial conditions in numerical simulations, stability analysis, and wave propagation studies, can lead to unphysical behavior. In particular, when studying wave propagation it is crucial to use exact equilibria to avoid spurious oscillations that can arise from the use of shifted Maxwellian distribution functions \cite{Malara2018}.

Numerous spacecraft measurements and observations have revealed that large amplitude, localized electrostatic structures known as Bernstein-Greene-Kruskal (BGK) modes are prevalent in space plasmas, where they manifest as solitary wave structures in both the magnetosphere and the solar wind \cite{Matsumoto1994,Ergun1998}. Theoretically, BGK modes arise as solutions to the Vlasov-Poisson system and were first proposed by Bernstein, Greene and Kruskal \cite{Bernstein1957}. Their existence has been verified through numerical simulations \cite{Demeio1991} and experimental observations \cite{Fox2008}. Although they are usually interpreted as electron phase space holes \cite{Hutchinson2017}, recent observations from NASA's Magnetospheric Multiscale (MMS) Mission \cite{Burch2016} have revealed the presence of a new type of bipolar electrostatic structures observed in a supercritical quasi-perpendicular Earth's bow shock crossing by the MMS. These structures have been interpreted as ion BGK modes  or ion phase space holes \cite{Wang2020,Vasko2018,Vasko2020}, as they are produced  by monopolar negative electrostatic potentials and their formation is attributed to the  ion two-stream instability  arising  from the interaction of incoming and reflected ions. In light of these observations, a theory of ion phase-space holes was recently proposed in \cite{Aravindakshan2020}, which builds upon earlier studies by \cite{Schamel1971,Schamel1986,Ng2005,Chen2004}. Within this framework the electrons are considered adiabatic and the dynamics of their distribution function is neglected. This is due to the fact that the velocities of ion BGK modes are significantly slower than the electron thermal velocity. However, like the standard theory of BGK modes, this approach is purely electrostatic and does not take into account the magnetic field or the interaction with a background plasma component. In the last two decades, there have been significant developments in the theory of BGK modes incorporating magnetic fields  \cite{Ng2020,Ng2006,Allanson2016b}), since the magnetic field plays an important role in the environments where BGK modes are observed. 

In this work, we adopt the abovementioned hybrid fluid-kinetic model to construct equilibria that take into account the self-consistent effects of  magnetic fields and can be interpreted as quasineutral BGK modes.  When dealing with wave structures characterized by large time and length scales, i.e., ones  much larger than the  electron time and the Debye length scale, there is sufficient time for the electrons to rearrange accordingly so that the hybrid approach and the quasineutrality condition are physically plausible. By employing this model we implicitly assume that the length scale of the wave is comparable to the ion inertial length $\ell_i$, i.e. much larger than the Debye length $\ell_D$. Thus the model is not well-suited for describing smaller bipolar structures such as the ion holes observed by the MMS mission during Earth's bow shock crossing, as these structures have spatial scales up to 10 $\ell_D$ \cite{Vasko2020}. In future work, we will investigate possible improvements to our model, that would allow for the construction of electrostatic structures with length scales $\sim \ell_D$ and nonzero charge separation.  For now, our focus is mainly on the impact of magnetic field variations and contributions stemming from fluid electrons on the formation of ion kinetic waves. Thus, we consider waves with characteristic lengths of approximately $\ell_i$ and neglect the self-consistent charge separation. A similar approach was employed in \cite{Tasso1969} to study the problem of electrostatic BGK modes, demonstrating that the quasineutrality condition ensures the positivity of the trapped particle distribution function. To construct stationary states,  we assume planar spatial symmetry and consider distribution functions that depend on the particle constants of motion, ensuring  that the Vlasov equation is satisfied. Then we solve an inhomogeneous Beltrami equation that arises from the generalized Ohm's law of the model, upon assuming that the electrons are described by Boltzmann distributions. This is in essentially an integro-differential equation that can be turned into a differential equation for the vector potential components if the form of the distribution function is specified, or conversely, into an integral equation for the distribution function upon defining the magnetic field. Within this framework, the electrostatic potential can be determined by the quasineutrality condition as a function of the vector potential components (e.g.\ see \cite{Mynick1979,Kuiroukidis2015,Tasso2014}). 

The electron contribution, given by $\lambda \textbf{B}$ in Ampere's law (an inhomogeneous Beltrami equation),  has some interesting ramifications. One such example is the modulation of the shape and distribution of the bipolar electric field pulses, which can be linked to spatial fluctuations in the magnetic field, caused by the electron current. A nonzero Beltrami parameter can lead to the emergence of multiple localized ion phase-space holes, even when the free parameters and initial conditions are set up to avoid such structures for $\lambda = 0$.  In addition, we have noticed that these structures exhibit periodicity for ion distribution functions that are symmetric in the two particle momentum constants of motion associated with the two ignorable coordinates. However, under suitable selection of initial conditions and for asymmetric distribution functions they appear in an irregular and chaotic manner. Both observations can be explained by examining the Hamiltonian phase-space structure of the dynamical system that governs the spatial evolution of the magnetic field vector potential components. This system originates from the generalized Beltrami-Ampere equation and the definition of the vector potential. More specifically, we show that these equations form a noncanonical Hamiltonian system (see e.g.\ \cite{Morrison1998}) that can be canonized under an appropriate transformation. The two-degree of freedom system possesses a first integral of motion, namely the Hamiltonian function. However, for symmetric ion distribution functions, there exists an additional integral, similar to the angular momentum, and the system is thus integrable. For asymmetric distribution functions though, this additional integral does not exist, and the orbits in parts of the phase-space become nonintegrable and chaotic. This is not surprising as it is well-established that magnetic field lines form  Hamiltonian systems \cite{Morrison2000} of more than one degree-of-freedom, which are typically chaotic. Furthermore, chaotic behavior is not unexpected in the context of kinetic equilibria. Previous studies, such as \cite{Ghosh2014}, have reported similar behavior of magnetic fields in Vlasov-Maxwell equilibria. However, our analysis is more general than that of \cite{Ghosh2014} because we prove the Hamiltonian nature of the equilibrium system for a wider class of distribution functions. Also, our analysis is concerned with the hybrid Vlasov-Maxwell model that treats the electrons as a massless fluid. In terms of equilibria,  this translates into an electron current density of the form $\lambda \textbf{B}$. The electron contribution gives rise to Coriolis-like (gyroscopic) coupling terms in the Hamiltonian system, which alter significantly the dynamics of the system as illustrated by various Poincar\'e surfaces of section. An interesting modification induced by the electron current is that unbounded trajectories of the Hamiltonian system can be confined by considering nonzero values of $\lambda$. 

The rest of the paper is structured as follows. In Section \ref{sec_II}, we derive the quasineutral hybrid Vlasov model from a parent non-quasineutral model using appropriate ordering. Section \ref{sec_III} describes the general method for constructing equilibrium solutions, further specializing to the one-dimensional (1-D) case. In Section \ref{sec_IV}, we demonstrate that the equations governing the equilibrium magnetic field form a noncanonical Hamiltonian system that can be transformed into a canonical one under an appropriate change of variables. In Section \ref{sec_V}, we select a particular distribution function and numerically integrate the magnetic field equilibrium equations and we use Poincar\'e surfaces of section to study the phase-space structure. In Section \ref{sec_VI}, we provide a paradigm for constructing analytic equilibria by specifying the pseudopotential associated with the Hamiltonian system, rather than the distribution function. The distribution function is subsequently reconstructed using a Hermite polynomial expansion method. Finally, in Section \ref{sec_VII}, we summarize the results and propose future extensions of the present study.

\section{The hybrid model and equilibrium formulation}
\label{sec_II}

In the study of slow electrostatic structures like ion BGK modes, which propagate with the ion acoustic velocity, electron dynamics can be neglected and the electrons can be treated adiabatically. This is possible because the electron thermal velocities are typically much greater than the ion acoustic velocity, making electron particle-wave resonances and other kinetic effects  irrelevant. To study such cases we can start our analysis with a hybrid Maxwell-Vlasov system that accounts for massless electrons and allows for charge separation:
\begin{eqnarray}
&&\partial_t f+\bsv\cdot\nb f+\frac{e}{m}(\bfE+\bsv\times \bfB)\cdot\nb_{\bsv} f =0\,,\label{vlasov_1}\\
&&\textbf{E}= -\textbf{u}\times \textbf{B}+\frac{1}{en_e}\textbf{J}\times \textbf{B}-\frac{\nb P_e}{en_e}\,, \label{Ohm_1}\\
&&\partial_t\bfB=-\nb\times\bfE\,,\quad \nb\cdot\bfB=0\,, \label{maxwell_1}\\
&&\mu_0\epsilon_0\partial_t\bfE=\nb\times\bfB-\mu_0\bfJ\,, \label{ampere_1}\\
&&\nb\cdot\bfE=\epsilon_0^{-1}\sigma\,, \label{gauss_1}\\
&& P_e=n_ek_B T_{e}\,. \label{eos_1}
\end{eqnarray}
where $f$ is the ion distribution function, $\sigma$ is the charge density given by
\begin{eqnarray}
\sigma=e\left(\int d^3v f -n_e\right)\,,
\end{eqnarray}
 the current density is
\begin{eqnarray}
\textbf{J}=e\left(\int d^3v\, \bsv f -n_e \textbf{u}_e\right)= \textbf{J}_k- en_e\textbf{u}_e\,,
\end{eqnarray}
where the kinetic ion current is
\begin{eqnarray}
\textbf{J}_k=en\textbf{u}=e\int d^3v\, \bsv f\,, \label{u_J_ion}
\end{eqnarray}
and other symbols are standard.
To close the system we assume an isothermal electron fluid, i.e.,   $T_e=T_{e0}=const.$. The Gauss law \eqref{gauss_1} can be used for determining the electron density $n_e$. 

Stationary solutions of the system \eqref{vlasov_1}--\eqref{eos_1}, can be found upon setting $\partial/\partial_t =0$, while travelling wave structures can be studied by a Galilean transformation from the laboratory frame.  Thus, we consider  the following system of equilibrium equations:
\begin{eqnarray}
&&\bsv\cdot \nb f + \frac{e}{m}\left(\textbf{E}+\bsv\times \textbf{B}\right)\cdot \nb_{\bsv} f=0\,,
 \label{vlasov_2}\\
&&\textbf{E}= -\textbf{u}\times \textbf{B}+\frac{1}{en_e}\textbf{J}\times \textbf{B}-\frac{\nb P_e}{en_e}\,, \label{ohm_2}\\
&&\textbf{E}=-\nabla\Phi\,, 
\quad \nb\times \textbf{B}=\mu_0\textbf{J}\,,\quad \nb\cdot\textbf{B}=0\,, \label{maxwell_2}\\
&& \nabla\cdot\textbf{E}=e\left(\int d^3v\, f-n_e\right)\,, \label{gauss_2}\\
&& P_e=n_e k_B T_{e0}\,. \label{eos_2}
\end{eqnarray}
The system \eqref{vlasov_2}--\eqref{eos_2} can describe a wide variety of physical processes ranging from ion BGK modes to nonlinear Alfv\'en waves. To see this let us consider two different normalization schemes. The first scheme involves a normalization in terms of microscopic characteristic quantities, such as the ion Debye length, introducing the following nondimensional quantities:
\begin{eqnarray}
&&\tilde{x}=\frac{x}{\ell_{D}}\,, 
\quad \tilde{\bsv}=\frac{\bsv}{v_{th,i}} \,, \nn \\
&&\tilde{n}_e=\frac{n_e}{n_0}\,, \quad
\tilde{f}=v_{th,i}^3 f/n_0\,, \nn \\
&&\tilde{\textbf{E}}=e\ell_{D}\frac{\textbf{E}}{k_B T_i}\,, \quad \tilde{\textbf{B}}=\frac{\textbf{B}}{ \mu_0en_0\ell_{D}v_{th,i}}\,,\nn\\
&&\tilde{\textbf{J}}_k=\frac{\textbf{J}_k}{en_0 v_{th,i}}\,, \quad \tilde{P}_e=\frac{P_e}{k_B n_0 T_{e0}}\,,
\end{eqnarray}
where $n_0$ is a characteristic background density, $T_{i0}$ is a characteristic ion temperature and 
\begin{eqnarray}
\ell_D=\sqrt{\frac{\epsilon_0 k_B T_i }{n_0 e^2}}\,,\quad v_{th,i}=\sqrt{\frac{k_B T_i}{m}}\,,
\end{eqnarray}
are the ion Debye length and the ion thermal velocity, respectively.
In terms of these nondimensional quantities, upon dropping the tildes,  Eqs.~\eqref{vlasov_2}--\eqref{eos_2} become
\begin{eqnarray}
&&\bsv\cdot\nabla f+\left(\textbf{E}+\beta_{th,i}^2\bsv\times\textbf{B} \right)\cdot \nabla_{v}f =0\,, \label{vlasov_3} \\
&&\textbf{E} = \frac{\beta_{th,i}^2}{n_e}\left(\nabla \times \textbf{B}-\textbf{J}_k\right)\times\textbf{B} - \nabla \ln n_e^\tau\,, \label{ohm_3}\\
&& \nabla \times \textbf{B} = \textbf{J}\,, \quad \nabla \cdot\textbf{B}=0\,,\\
&&\nabla\cdot\textbf{E} = \int d^3v\, f - n_e\,, \label{gauss_3}
\end{eqnarray}
where $\tau = T_{e0}/T_{i0}$, $\beta_{th,i}^2=v_{th,i}^2/c^2$ and $P_e=n_e$. 

In the nonrelativistic limit where $\beta_{th,i}^2=v_{th,i}^2/c^2 \rightarrow 0$ the system \eqref{vlasov_3}--\eqref{gauss_3} reduces to the Vlasov-Poisson system for the ions, while the electrons are adiabatic with their particle density described by the Boltzmann distribution
\begin{eqnarray}
n_e = exp(\Phi/\tau)\,, \label{ne_1}
\end{eqnarray} 
which has been used in previous studies for the description of ion holes  \cite{Aravindakshan2020} (note that upon  taking the limit $\kappa\rightarrow \infty$ in equation (5) of \cite{Aravindakshan2020}, one recovers \eqref{ne_1}). Therefore, with this particular ordering, magnetic field effects in the ion hole formation can be introduced upon keeping terms of the order $\beta_{th,i}^2$.  

Now, let us consider an alternative normalization, which is more suitable in cases where the magnetic fields are strong and charge separation is insignificant:  
\begin{eqnarray}
&&\tilde{x}=\frac{x}{\ell_i}\,, \quad \tilde{\bsv}=\frac{\bsv}{v_{A}} \,, \nn \\
&& \tilde{n}_e=\frac{n_e}{n_0}\,, \quad
\tilde{f}=\frac{v_{A}^3 f}{n_0}\,, \nn \\
&&\tilde{\textbf{E}}=\frac{\textbf{E}}{v_AB_0}\,, \quad \tilde{\textbf{B}}=\frac{\textbf{B}}{ B_0}\,,\nn\\
&&\tilde{\textbf{J}}_k=\frac{\textbf{J}_k}{en_0 v_{A}}\,, \quad \tilde{P}_e=\frac{P_e}{m n_0 v_A^2}\,,
\end{eqnarray}
where $B_0$ is a characteristic value of the magnetic field modulus and 
\begin{eqnarray}
\ell_i = v_A/\Omega\,,\quad v_A=\frac{B_0}{\sqrt{\mu_0 m n_0}}\,, \quad
\Omega &=& \frac{eB_0}{m}\,,
\end{eqnarray}
are the ion inertial length, the Alfv\'en velocity and the ion cyclotron frequency, respectively.

With this normalization, the system \eqref{vlasov_2}--\eqref{eos_2} reads as follows:
\begin{eqnarray}
&&\bsv\cdot\nb f+\left(\textbf{E}+\bsv \times \textbf{B}\right)\cdot \nb_{\bsv} f=0\,, 
\label{vlasov_4}\\
&&-\nabla \Phi=\left[\frac{1}{n_e} (\nb\times\textbf{B}-\textbf{J}_k)\times 
\textbf{B}- \nb \ln n_e^{\kappa} \right]\,,\label{ohm_4} \\
&&\nabla\times \textbf{B} =  \textbf{J}\,,\quad 
\nabla\cdot\textbf{B}=0\,, \label{max_4}\\ 
&&\beta_{A}^2 \nabla \cdot \textbf{E} = (n_i-n_e)\,.\label{gauss_4}
\end{eqnarray} 
In the above equations, $\kappa=k_B T_{e0}/(mv_A^2)$ and  $\beta_A^2:= v_A^2/c^2$. By using this ordering scheme, in the  nonrelativistic limit  where  $\beta_A^2 \rightarrow 0$, the Gauss law implies the quasineutrality condition $n_i=n_e$ and we obtain a system of equilibrium equations for the ordinary quasineutral hybrid Vlasov-Maxwell system (e.g.\  see \cite{Malara2018}). Here, when effects due to charge separation cannot be neglected, we need to keep terms of the order $\beta_A^2$.

\section{Equilibrium solutions}
\label{sec_III}
In this section we consider the system \eqref{vlasov_4}--\eqref{gauss_4} in the limit $\beta_A^2\rightarrow 0$. Within this framework we will calculate certain quasineutral, 1-D equilibria where all equilibrium quantities depend on a single spatial coordinate, denoted by $x$. To study the equilibrium problem one can in general exploit the pressure tensor formulation (e.g. see \cite{Mynick1979,Harrison2009,Allanson2016}). Specifically, we begin by considering the first velocity moment of the Vlasov equation \eqref{vlasov_4}, giving
\begin{eqnarray}
    \nabla \cdot \textbf{P} = n \textbf{E} + \textbf{J}_{k}\times \textbf{B}\,, \label{equil_eq_P_1}
\end{eqnarray}
where 
\begin{eqnarray}
    \textbf{P} = \int d^3v \, \bsv \bsv f\,,
\end{eqnarray}
is the ion pressure tensor. Upon inserting the generalized Ohm's law \eqref{ohm_4} into \eqref{equil_eq_P_1} and using the quasineutrality condition we find
\begin{eqnarray}
    (\nabla \times \textbf{B})\times \textbf{B} = \nabla \cdot \textbf{P}+ \kappa \nabla n\,. \label{equil_eq_P_2}
\end{eqnarray}
This equation can be seen either as a differential equation for determining the magnetic field $\textbf{B}$ upon specifying $f$, or as an integral equation to determine $f$ upon specifying $\textbf{B}$. 

For the rest of the paper we will consider 1-D equilbria assuming that  all physical quantities depend only on one cartesian coordinate $x$,  while the distribution function can depend on all three velocity coordinates $(v_x,v_y,v_z)$.  This consideration is physically relevant as various measurements of space plasmas have revealed that certain wave structures, e.g., BGK modes in the Earth's bow-shock (e.g. \cite{Wang2020,Vasko2018}) are nearly 1-D structures, in the sense that the electric field is highly polarized.  In this case the force-balance equation \eqref{equil_eq_P_2} yields
\begin{eqnarray}
    P_{xx} + \kappa n + \frac{B^2}{2} = V + \frac{B^2}{2} = const.\,, \label{pressure_balance}
\end{eqnarray}
where $P_{xx}$ is the $xx$ component of the ion pressure tensor and $n=n_i=n_e$:
\begin{eqnarray}
    P_{xx} =  \int d^3v\, v_x^2 f\,\\
    n= \int d^3v f\,.
\end{eqnarray}
Equation  \eqref{pressure_balance} is a pressure balance equation, where the first, second and third terms in the lhs of the first equality correspond to the ion, electron and magnetic pressure, respectively. Equation \eqref{pressure_balance} can be seen also as an expression for the energy conservation of a pseudoparticle with velocity components $B_y=-dA_z/dx$ and $B_z=dA_y/dx$ moving in a pseudopotential $V=P_{xx}+\kappa n$. Thus, the energy (Hamiltonian) of the pseudoparticle is $\Hc = B^2/2 + P_{xx}+\kappa n$.

An equilibrium state of a 1-D hybrid Vlasov system should satisfy \eqref{pressure_balance}. As the $y$ and $z$ coordinates are ignorable, the corresponding canonical momenta, 
\begin{eqnarray}
p_y = v_y+A_y\,,\\
p_z = v_z + A_z\,,
\end{eqnarray} 
are particle constants of motion of the kinetic ions, as is the energy $H=v^2/2+\Phi$. Note that the canonical momenta have been written in nondimensional form,  normalizing the vector potential as $\tilde{\textbf{A}}=\textbf{A}/(v_AB_0/\Omega)$, and that $H$ is different from $\Hc$. The former refers to the physical energy of the kinetic ions while the latter represents the energy of the pseudoparticle whose velocity corresponds to the equilibrium magnetic field. By Jean's theorem, any  function of $p_y$, $p_z$ and $H$ will solve the Vlasov equation \eqref{vlasov_5}. This can be easily verified by substituting $f=f(H,p_y,p_z)$ in the Vlasov equation expressed in Cartesian spatial and velocity coordinates. 

The final step to fully define the equilibrium problem is to introduce a relation that connects the electron density with the electrostatic potential. This is usually done through the quasineutrality condition. Although we have used quasineutrality to replace the electron density $n_e$ with the ion density $n_i=n$, the equation $n_i=n_e$ cannot be used to determine $\Phi$ in terms of $A_y$, $A_z$, since there is no information about how the number density of the electrons is related to $\Phi$, $A_y$ and $A_z$. To overcome this problem we can define quasineutrality in a way similar to \cite{Mynick1979}, i.e.
\begin{eqnarray}
    \frac{\partial V}{\partial \Phi} = \kappa \frac{\partial n}{\partial \Phi} + \frac{\partial P_{xx}}{\partial \Phi}= 0\,. \label{quasineutrality}
\end{eqnarray}
We can show, however, (see \cite{Harrison2009}) that $\partial P_{xx}/\partial \Phi = -n$ and therefore \eqref{quasineutrality} reads
\begin{eqnarray}
    \kappa \frac{\partial n}{\partial \Phi} =n\,,
\end{eqnarray}
which is satisfied by number density functions of the form 
\begin{eqnarray}
    n = e^{\Phi/\kappa}W(A_y,A_z)\,,
\end{eqnarray}
where $W$ is an arbitrary function of $A_y$ and $A_z$. If we require now the components $J_{ky}$ and $J_{kz}$ of the kinetic current density to be given by partial derivatives of the pseudopotential $V$ with respect to $A_y$ and $A_z$, respectively, we have
\begin{eqnarray}
    J_{ky} = \frac{\partial P_{xx}}{\partial A_y} +\kappa e^{\Phi/\kappa} \frac{\partial W}{\partial A_y}= J_{ky} + \kappa e^{\Phi/\kappa} \frac{\partial W}{\partial A_y}\,, \label{partial_yW}\\
    J_{kz} = \frac{\partial P_{xx}}{\partial A_z} + \kappa e^{\Phi/\kappa} \frac{\partial W}{\partial A_z} = J_{kz} +\kappa e^{\Phi/\kappa} \frac{\partial W}{\partial A_z}\,,\label{partial_zW}
\end{eqnarray}
where we have used $\partial P_{xx}/\partial A_y = J_{ky}$ and $\partial P_{xx}/\partial A_z = J_{kz}$ (\cite{Mynick1979,Harrison2009}). Equations \eqref{partial_yW} and  \eqref{partial_zW} imply $W(A_y,A_z) =const.$ and therefore the number density is described by a Boltzmann distribution of the form
\begin{eqnarray}
n = exp(\Phi/\kappa)\,. \label{ne_2}
\end{eqnarray}
Note that in the calculations above we consider $\Phi$, $A_y$ and $A_z$ as being independent,  although $\Phi$ can be expressed in terms of $A_y$ and $A_z$ through $e^{\Phi/\kappa}W(A_y,A_z)=\int d^3v f(H,p_y,p_z)$, as  will be the case in the next section.

With \eqref{ne_2} we can readily see that the system \eqref{vlasov_4}--\eqref{gauss_4} splits into
\begin{eqnarray}
\bsv\cdot\nb f+\left(-\nb \Phi+\bsv \times \textbf{B}\right)\cdot\nb_{\bsv}f=0\,,\label{vlasov_5}\\
\nb \times \textbf{B}=\lambda \textbf{B} +\textbf{J}_k\,. \label{beltrami}
\end{eqnarray}
This splitting implies that the electron current density is $\lambda \textbf{B}$, i.e.,  it is aligned with the magnetic field. This is justified by the fact that inertialess electrons move along the magnetic field lines. Note that in general the coefficient $\lambda$, which will be  called the  Beltrami or Coriolis parameter, may be a function of some spatial coordinate. In this case though, the following equations should be satisfied by $\lambda$:
\begin{eqnarray}
\textbf{B}\cdot\nabla \lambda(\textbf{x}) + \nabla \cdot \textbf{J}_k=0\,,
\end{eqnarray}
or 
\begin{eqnarray}
    \textbf{B} \cdot \nabla \lambda(\textbf{x}) =0\,,
\end{eqnarray}
since  $\nabla\cdot \textbf{J}_k =0$ at equilibrium. In the 1-D case, \eqref{beltrami} reduces to a system of two integro-differential equations for determining $A_y$ and $A_z$,  These equations are
\begin{eqnarray}
\frac{d^2 A_y}{dx^2}-\lambda \frac{d A_z}{dx}+ J_{ky}=0\,, \label{beltrami_plan_1}\\
\frac{d^2 A_z}{dx^2}+\lambda \frac{d A_y}{dx}+ J_{kz}=0\,, \label{beltrami_plan_2}
\end{eqnarray}
where
\begin{eqnarray}
J_{ky} = \int d^3v f v_y\,,\\
J_{kz} = \int d^3v f v_z\,,
\end{eqnarray}
and $\lambda$ is a constant parameter. If the distribution function is a function of the form $f=f(H,p_y,p_z)$, the Vlasov equation and consequently \eqref{pressure_balance} are satisfied automatically. 

\section{Equilibrium equations as a Hamiltonian system}
\label{sec_IV}
\subsection{Hamiltonian structure}
\label{subsec_IV_i}
Since the 1-D equilibrium magnetic field can be represented by the velocity of a pseudoparticle with conserved energy, it is natural to seek for a Hamiltonian structure in the equilibrium equations. The second-order system composed of  \eqref{beltrami_plan_1}--\eqref{beltrami_plan_2} can be reduced to a first-order system of ordinary differential equations using $\textbf{B}=\nabla\times \textbf{A}$, viz.  
\begin{eqnarray}
&&\frac{dA_y}{dx}=B_z\,,\label{dAy_dx}\\
&&\frac{dA_z}{dx}=-B_y\,,\label{dAz_dx}\\
&&\frac{dB_y}{dx}=\lambda B_z + J_{kz}\,,\label{dBy_dx}\\
&&\frac{dB_z}{dx}=-\lambda B_y - J_{ky}\,.\label{dBz_dx}
\end{eqnarray}
Let us consider distribution functions of the special form
\begin{eqnarray}
f=f(H,p_y,p_z) = exp(-H)g(p_y,p_z)\,, \label{DF_gen_form}
\end{eqnarray}
where $g$ is an arbitrary but well-behaved and positive function of the two canonical momenta. With this choice, the particle and current densities are given by
\begin{eqnarray}
\hspace{-10mm}n_i &=& \int d^3v\,  e^{-H}g(p_y,p_z)=\int_{-\infty}^\infty \!\!dp_y \int_{-\infty}^\infty \!\!dp_z  \!\int_{H^*}^{\infty}\!\!dH\, \frac{e^{-H}g(p_y,p_z)}{\sqrt{2(H-H^*)}}\,,\\
\hspace{-10mm}J_{ky} &=& \int\! d^3v \, v_y e^{-H}g(p_y,p_z)= \int_{-\infty}^\infty \!\!dp_y \int_{-\infty}^\infty\!\!  dp_z \!\int_{H^*}^{\infty}\!\! dH\, (p_y-A_y) \frac{e^{-H}g(p_y,p_z)}{\sqrt{2(H-H^*)}}\,,\\
\hspace{-10mm}J_{kz} &=& \int\! d^3v \, v_z e^{-H}g(p_y,p_z)= \int_{-\infty}^\infty \!\!dp_y \int_{-\infty}^\infty \!\! dp_z \! \int_{H^*}^{\infty}\!\! dH\, (p_z-A_z) \frac{e^{-H}g(p_y,p_z)}{\sqrt{2(H-H^*)}}\,,
\end{eqnarray}
where
\begin{eqnarray}
    H^* = \Phi + \frac{(p_y-A_y)^2}{2} + \frac{(p_z-A_z)^2}{2}\,.
\end{eqnarray}
Integration with respect to $H$ yields
\begin{eqnarray}
\hspace{-10mm}n_i &=& \sqrt{\frac{\pi}{2}} e^{-\Phi} \int_{-\infty}^\infty \!\! dp_y \int_{-\infty}^\infty \!\! 
dp_z \, e^{-\frac{1}{2}(p_y-A_y)^2-\frac{1}{2}(p_z-A_z)^2}g(p_y,p_z)\,, 
\label{ni_gen}\\
\hspace{-10mm}J_{ky} &=& \sqrt{\frac{\pi}{2}} e^{-\Phi} \int_{-\infty}^\infty \!\! dp_y\int_{-\infty}^\infty \!\! dp_z \, (p_y-A_y)e^{-\frac{1}{2}(p_y-A_y)^2-\frac{1}{2}(p_z-A_z)^2}g(p_y,p_z)\,,
 \label{Jy_gen}\\
\hspace{-10mm}J_{kz} &=& \sqrt{\frac{\pi}{2}} e^{-\Phi} \int_{-\infty}^\infty \!\! dp_y\int_{-\infty}^\infty \!\! dp_z \, (p_z-A_z)e^{-\frac{1}{2}(p_y-A_y)^2-\frac{1}{2}(p_z-A_z)^2}g(p_y,p_z)\,.
\label{Jz_gen}
\end{eqnarray}

At this point we use the quasineutrality condition $n_i=n_e$ rather than assuming $\Phi=0$ which was the case in other studies (e.g.\  \cite{Ghosh2014,Allanson2016}). Upon imposing quasineutrality and using \eqref{ne_2}  we have
\begin{eqnarray}
n_i=n_e=exp(\Phi/\kappa)\,.
\end{eqnarray}
Using \eqref{ni_gen} and solving for the electrostatic potential $\Phi$ yields
\begin{eqnarray}
\Phi=\Phi(A_y,A_z;\kappa)=\frac{\kappa}{\kappa +1} ln \big(G(A_y,A_z)\big)\,, \label{Phi_qn}
\end{eqnarray}
where 
\begin{eqnarray}
G(A_y,A_z) = \sqrt{\frac{\pi}{2}}\int_{-\infty}^{\infty} \!\! dp_y \int_{-\infty}^{\infty}  \!\! dp_z \, e^{-\frac{1}{2}(p_y-A_y)^2-\frac{1}{2}(p_z-A_z)^2}g(p_y,p_z)\,. 
\label{G_fun}
\end{eqnarray}
Upon inserting the expression \eqref{Phi_qn} for the electrostatic potential into Eqs.~\eqref{Jy_gen}--\eqref{Jz_gen},  we can show that
\begin{eqnarray}
J_{ky}= (\kappa+1)\, \frac{\partial G^{\frac{1}{\kappa+1}}}{\partial A_y}\,,\quad
J_{kz}= (\kappa+1)\, \frac{\partial G^{\frac{1}{\kappa+1}}}{\partial A_z}\,, 
\end{eqnarray}
that is, the two components of the current density can be derived from the function
\begin{eqnarray}
V(A_y,A_z) = (\kappa+1)\, G^{\frac{1}{\kappa+1}} \,,
\label{V_gen}
\end{eqnarray}
as
\begin{eqnarray}
J_{ky}=\frac{\partial V}{\partial A_y}\,, \quad J_{kz}=\frac{\partial V}{\partial A_z}\,. \label{J_from_V}
\end{eqnarray}
This same result was  obtained in \cite{Channell1976,Mynick1979} for $\Phi=0$ and $\Phi\neq 0$. Note that the pseudopotential $V$ is in fact $V=(\kappa+1) \tilde{n} =(\kappa+1)\tilde{P}_{xx}$, where $\tilde{n}=n(\Phi(A_y,A_z),A_y,A_z)$ and $\tilde{P}_{xx}=P_{xx}(\Phi(A_y,A_z),A_y,A_z)$. Hence, \eqref{pressure_balance} yields
\begin{eqnarray}
    V(A_y,A_z) + \frac{B^2}{2} = \Hc=const\,. \label{Hamiltonian_1}
\end{eqnarray}

Now, for $V$ to be a valid potential function it should satisfy
\begin{eqnarray}
\frac{\partial ^2 V}{\partial A_y \partial A_z} = \frac{\partial^2 V}{\partial A_z \partial A_y}\,,
\end{eqnarray}
which can easily be verified from \eqref{V_gen}. Therefore, it can be seen that the dynamical system \eqref{dAy_dx}--\eqref{dBz_dx} is a noncanonical Hamiltonian system, i.e.,  it is of the form
\begin{eqnarray}
\dot{\boldsymbol{\xi}} = \boldsymbol{\Jc}\cdot \frac{\partial \Hc}{\partial \boldsymbol{\xi}}\,,
\end{eqnarray}
where $\boldsymbol{\xi} = (A_y,A_z,B_z,-B_y)$, $\Jc$ is a $4\times 4$ noncanonical Poisson matrix of the form
\begin{eqnarray}
\boldsymbol{\Jc}=
\begin{pmatrix}
0_2 && I_2 \\
-I_2 && \Lambda \\
\end{pmatrix}\,,  \qquad 
\Lambda = \begin{pmatrix}
0 && \lambda \\
-\lambda && 0
\end{pmatrix}\,,
\end{eqnarray}
where $0_2$ and $I_2$ are the $2\times 2$ zero and identity matrices, respectively, and the Hamiltonian function is given by  \eqref{Hamiltonian_1}.   Note that in the Hamiltonian framework, the conservation of $\Hc$ is a consequence of the antisymmetry of the Poisson matrix $\boldsymbol{\Jc}$. Also, the structure of the Poisson matrix implies that the Hamiltonian system can be canonized by the following change of variables $\boldsymbol{\xi}=(A_y,A_z,B_z,-B_y)\rightarrow \boldsymbol{\xi}_c=(Q_1,Q_2,P_1,P_2)$, where
\begin{eqnarray}
Q_1 = A_y\,, \quad Q_2 = A_z \,,\nonumber\\
P_1 = B_z-\frac{\lambda}{2} Q_2 = \dot{Q}_1-\frac{\lambda}{2} Q_2 \,,\nonumber \\
P_2 = -B_y+\frac{\lambda}{2}Q_1=-\dot{Q}_2+\frac{\lambda}{2}Q_1\,. \label{canonical_vars}
\end{eqnarray}
When expressed in  the  variables $\boldsymbol{\xi}_c$, the system \eqref{dAy_dx}--\eqref{dBz_dx} takes the form
\begin{eqnarray}
\dot{\boldsymbol{\xi}}_c= \boldsymbol{\Jc}_c\cdot\frac{\partial \bar{\Hc}}{\partial\boldsymbol{\xi}_c}\,,
\end{eqnarray} 
where
\begin{eqnarray}
\boldsymbol{\Jc}_c=\begin{pmatrix}
0_2 && I_2\\
-I_2 && 0_2
\end{pmatrix}\,,
\end{eqnarray}
and
\begin{align}
&\bar{\Hc}=\frac{1}{2}\left(P_1+\frac{\lambda}{2} Q_2\right)^2+\frac{1}{2}\left(P_2-\frac{\lambda}{2}Q_1\right)^2+V(Q_1,Q_2)\nn\\
&= \frac{1}{2}(P_1^2+P_2^2)+\frac{\lambda^2}{8}(Q_1^2+Q_2^2)+\frac{\lambda}{2}(P_1Q_2-P_2Q_1)+V(Q_1,Q_2)\,. \label{Hamiltonian_2}
\end{align}
We observe that for $\lambda\neq 0$, the Hamiltonian of the canonical system possesses two Coriolis-like  coupling terms that  would be absent if the contribution of the electrons in the current density  vanished. Those terms, would also be absent if we had considered the solely  kinetic Vlasov-Maxwell system rather than the hybrid variant with fluid, massless electrons. Such Coriolis couplings arise  in Hamiltonian models of finite Larmor  radius stabilization \cite{pjmK90}, plasma streaming instabilities \cite{pjmK95},  fluid models that describe collisionless magnetic reconnection \cite{pjmTWG08} and galactic dynamics \cite{Binney2008,Salas2022},  and play a significant role in the dynamical evolution of the relevant physical systems. The ramifications of this coupling, in terms of the phase space structure and escape dynamics in the context of our study are discussed in the next section.

In summary, the canonical Hamiltonian equations of motion read as follows:
\begin{eqnarray}
\dot{Q}_1&=& \frac{\partial \bar{\Hc}}{\partial P_1}= P_1+\frac{\lambda}{2} Q_2\,, \label{can_ham_eq_1}\\
\dot{Q}_2 &=& \frac{\partial \bar{\Hc}}{\partial P_2}=P_2-\frac{\lambda}{2}Q_1\,,\\
\dot{P}_1 &=& -\frac{\partial \bar{\Hc}}{\partial Q_1}= \frac{\lambda}{2}(P_2-\frac{\lambda}{2}Q_1)-\frac{\partial V}{\partial Q_1}\,,\\
\dot{P}_2 &=&-\frac{\partial \bar{\Hc}}{\partial P_2}=  -\frac{\lambda}{2}(P_1+\frac{\lambda}{2}Q_2)-\frac{\partial V}{\partial Q_2}\,. \label{can_ham_eq_4}
\end{eqnarray}

\section{Specifying the  distribution function: the forward equilibrium problem}
\label{sec_V}
In the direct or forward equilibrium problem, \eqref{V_gen} can be seen as an equation for determining the pseudopotential $V$ for some specific ion distribution of the form $f=exp(-H)g(p_y,p_z)$, and then the magnetic field and the electrostatic potential can be computed upon integrating the system \eqref{can_ham_eq_1}--\eqref{can_ham_eq_4} and invoking \eqref{Phi_qn} for $\Phi$. Here, we implement this procedure considering a special  distribution function  of the form \cite{Ng2020}
\begin{eqnarray}
f(H,p_y,p_z)=2^{-1/2}\pi^{-3/2}e^{-H}\left[1-\delta_1 exp\left(-\delta_2 p_y^2-\delta_3 p_z^2\right)\right]\,, \label{DF_1}
\end{eqnarray}
so that for $\delta_1<1$ the distribution function is always positive. In this case, using  \eqref{ni_gen}--\eqref{G_fun} and \eqref{V_gen} we obtain 
 
\begin{align}
&n = e^{-\Phi}\left[1-\gamma \Ec(Q_1,Q_2)\right]\,,
\label{ni_sp}\\
&J_{ky} = \frac{2\gamma \delta_2}{1+2\delta_2} Q_1 e^{-\Phi} \Ec(Q_1,Q_2)\,, \label{Jky_sp}\\
&J_{kz} = \frac{2 \gamma \delta_3}{1+2\delta_3} Q_2 e^{-\Phi} \Ec(Q_1,Q_2)\,,
\label{Jkz_sp} \\
&\Phi =  \frac{\kappa}{1+\kappa}\ln\left[1-\gamma \Ec(Q_1,Q_2)\right]\,, \label{Phi_sp}\\
&V(Q_1,Q_2) = (1+\kappa) \left[1- \gamma \Ec(Q_1,Q_2)\right]^{\frac{1}{k+1}} \,,\label{V_sp}
\end{align} 
where 
\begin{eqnarray}
\gamma = \frac{\delta_1}{\sqrt{1+2\delta_2}\sqrt{1+2\delta_3}}\,, \quad \Ec(Q_1,Q_2)=exp\left(-\frac{\delta_2 Q_1^2}{1+2\delta_2}-\frac{\delta_3  Q_2^2}{1+2\delta_3 }\right)\,.
\end{eqnarray}
For the particular pseudopotential function \eqref{V_sp}, the  Hamiltonian \eqref{Hamiltonian_2} becomes
\begin{eqnarray}
\bar{\Hc}&=& \frac{1}{2}\left(P_1^2 +P_2^2 \right) + \frac{\lambda^2}{8}\left(Q_1^2+Q_2^2\right)\nn\\
&&
-\frac{\lambda}{2}\left(  Q_1P_2 - Q_2 P_1\right) 
+ (1+\kappa) \left[1- \gamma e^{-\frac{\delta_2 Q_1^2}{1+2\delta_2} -\frac{\delta_3 Q_2^2}{1+2\delta_3}}\right]^{\frac{1}{k+1}}\,. 
\label{Hamiltonian_3}
\end{eqnarray}
Upon dropping  the overbar to avoid clutter, we observe the  Hamiltonian \eqref{Hamiltonian_3} possesses the following special discrete symmetries:
\begin{eqnarray}
\Sc_1 : \quad \Hc(Q_1,Q_2,P_1,P_2; \lambda) = \Hc(-Q_1,-Q_2,P_1,P_2; -\lambda)\,, \label{symmetry_1}\\
\Sc_2 : \quad \Hc(Q_1,Q_2,P_1,P_2; \lambda) = \Hc(Q_1,Q_2,-P_1,-P_2; -\lambda)\,, \label{symmetry_2}\\
\Sc_3 : \quad \Hc(Q_1,Q_2,P_1,P_2; \lambda) = \Hc(-Q_1,-Q_2,-P_1,-P_2; \lambda)\,, \label{symmetry_3}
\end{eqnarray}
which imply that we can  consider either  positive or negative Beltrami parameters $\lambda$.

\subsection{Influence of $\lambda$ on the linear stability}
\begin{figure}
\centering
\includegraphics[scale=0.45]{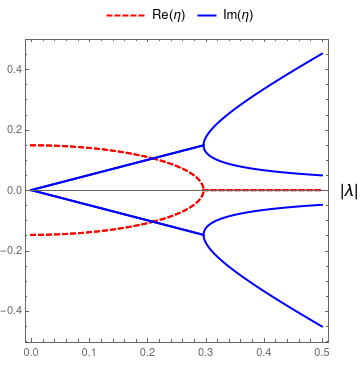}
\caption{Stabilization of the equilibrium point $\xi_e$ in the case $\delta_1<0$. Increasing $|\lambda|$, the stability of $\xi_e$ changes through a single Krein bifurcation for each eigenpair in the isotropic case ($\delta_2 = \delta_3$).}
\label{fig_krein_1}
\end{figure}

It can be readily seen that  $\xi_e:=(Q_1=0,Q_2=0, P_1=0,  P_2=0)$ is an equilibrium point of the dynamical system \eqref{can_ham_eq_1}--\eqref{can_ham_eq_4} with $V$ given by \eqref{V_sp}; it corresponds to an extremal of the Hamiltonian $ \Hc$ and a  global extremum of the potential $V$. This  is  a minimum of $V$ for $0<\gamma<1$  and maximum for $\gamma<0$. Linear stability analysis for the equilibrium point  reveals a stabilizing effect of the electron current density contribution manifested through a nonzero Beltrami parameter $\lambda$. The spectrum of the fixed point $\xi_e=(0,0,0,0)$ consists of two eigenpairs given  by (see Appendix \ref{app_1}):
\begin{eqnarray}
    \eta_{i} = \pm \frac{1}{\sqrt{2}} \sqrt{-\lambda^2-V_1-V_2 \pm \sqrt{(\lambda^2+V_1+V_2)^2-4V_1V_2}}\,, \quad i =1,...,4\,,
\end{eqnarray}
where
\begin{eqnarray}
V_1=\frac{2 \gamma\delta_2 (1-\gamma)^{\frac{-\kappa}{\kappa+1}}}{1+2\delta_2}\,, \quad
V_2 = \frac{2 \gamma\delta_3 (1-\gamma)^{\frac{-\kappa}{\kappa+1}}}{1+2\delta_3}\,, \label{V_1-V_2_nonlin}
\end{eqnarray}
For $0<\gamma<1$ the eigenvalues comprise two purely imaginary complex conjugate pairs, thus the equilibrium point is spectrally stable. Also in this case the Dirichlet criterion is satisfied (see e.g.\ \cite{Morrison1998}), i.e. the Hessian $D^2\Hc(\xi_e)$ is positive definite. On the other hand, if $\delta_1<0$, i.e. $\gamma<0$, then the stability of $\xi_e$ is determined by the value of $\lambda$. It can be readily verified that for $\lambda=0$ the eigenvalues comprise real pairs and thus $\xi_e$ is an unstable equilibrium point. However, as $|\lambda|$ increases the stability of the equilibrium point changes. In the symmetric case $\delta_2=\delta_3$ this happens through a single bifurcation;  the two pairs of complex conjugate eigenvalues bifurcate to two purely imaginary conjugate eigenpairs and thus the equilibrium point becomes stable (Fig. \ref{fig_krein_1}). This is the so-called Krein's bifurcation that occurs in the presence of negative energy modes (see, e.g.,  \cite{moser58}). 

\begin{figure}
\centering
\includegraphics[scale=0.45]{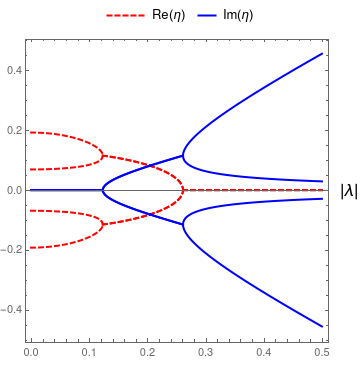}
\caption{Stabilization of the equilibrium point $\xi_e$ in the case $\delta_1<0$. Increasing $|\lambda|$, the stability of $\xi_e$ changes through two consecutive Krein bifurcations for each eigenpair in the anisotropic case ($\delta_2 \neq \delta_3$).}
\label{fig_krein_2}
\end{figure}

In the asymmetric case $\delta_2\neq\delta_3$ two consecutive bifurcations take place as $|\lambda|$ increases. First, the two pairs of real eigenvalues collide on the real axis and move to the complex plane forming two complex conjugate eigenpairs. These pairs subsequently crash on the imaginary axis bifurcating into two purely imaginary eigenpairs (Fig. \ref{fig_krein_2}). A normal-form Hamiltonian can also be written for this case.

\subsection{Influence of $\lambda$ on the escape dynamics and phase space structure}
Next  we examine the influence of the Beltrami parameter $\lambda$ on the escape dynamics using Poincare surfaces of section and discuss about its influence on the formation of bipolar electrostatic structures due to magnetic field fluctuations.

\begin{figure}
\centering
\includegraphics[scale=0.135]{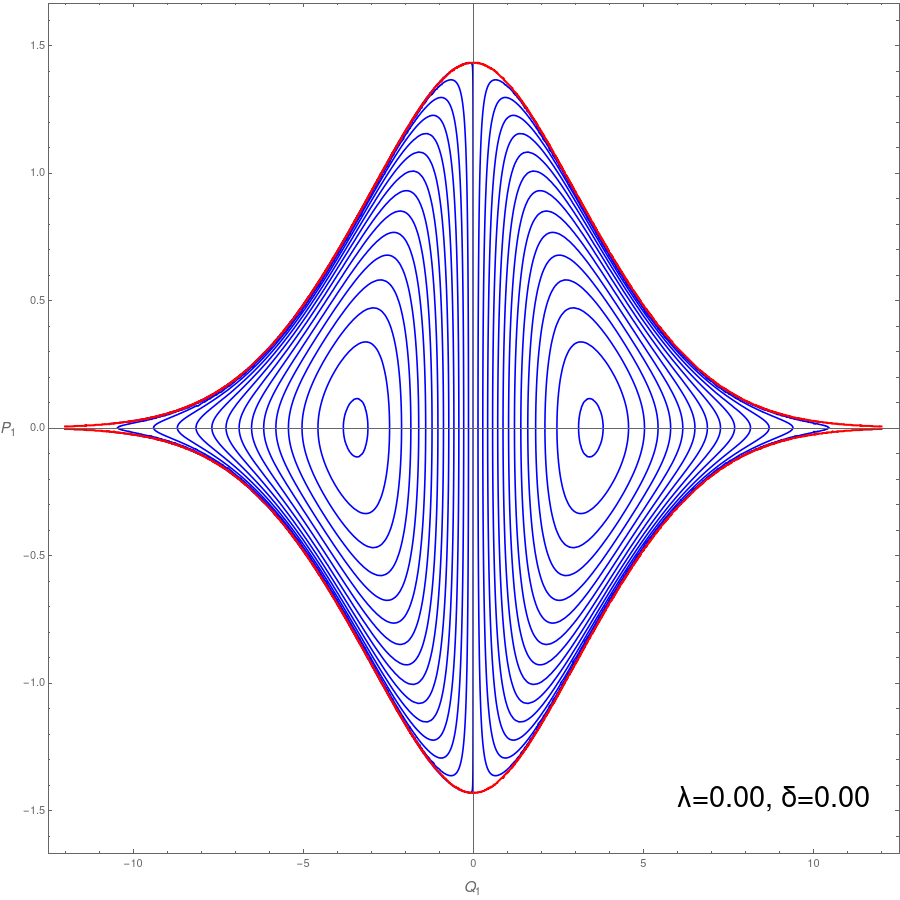}
\includegraphics[scale=0.135]{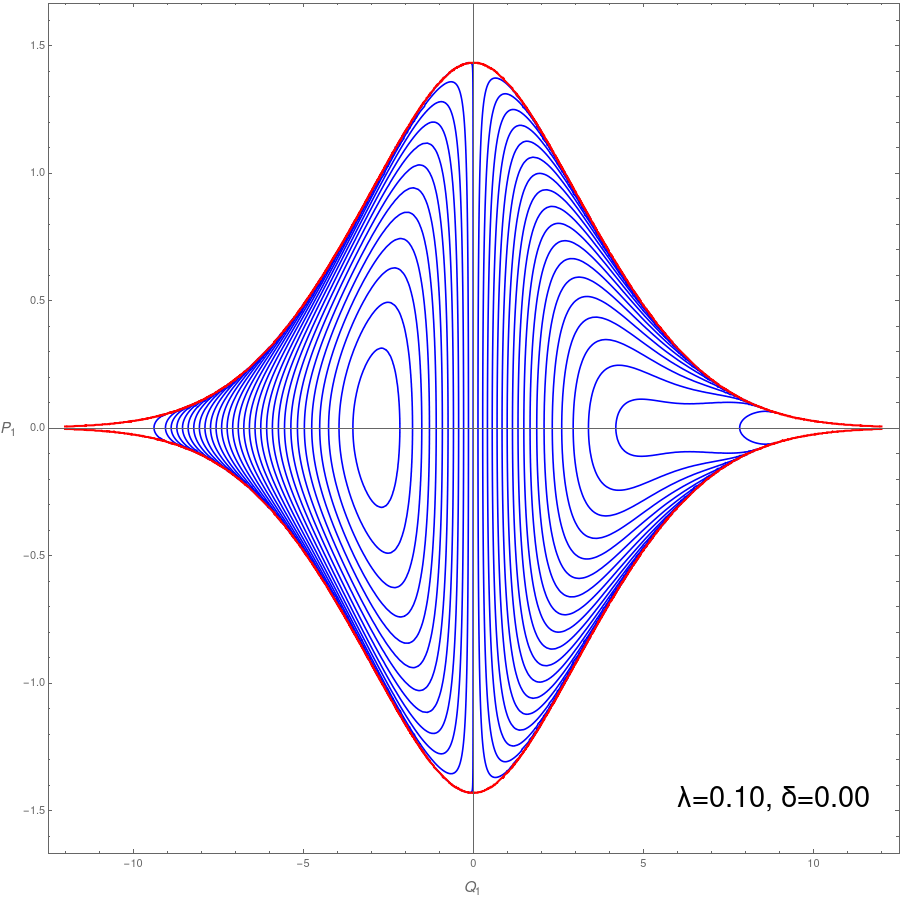}
\includegraphics[scale=0.135]{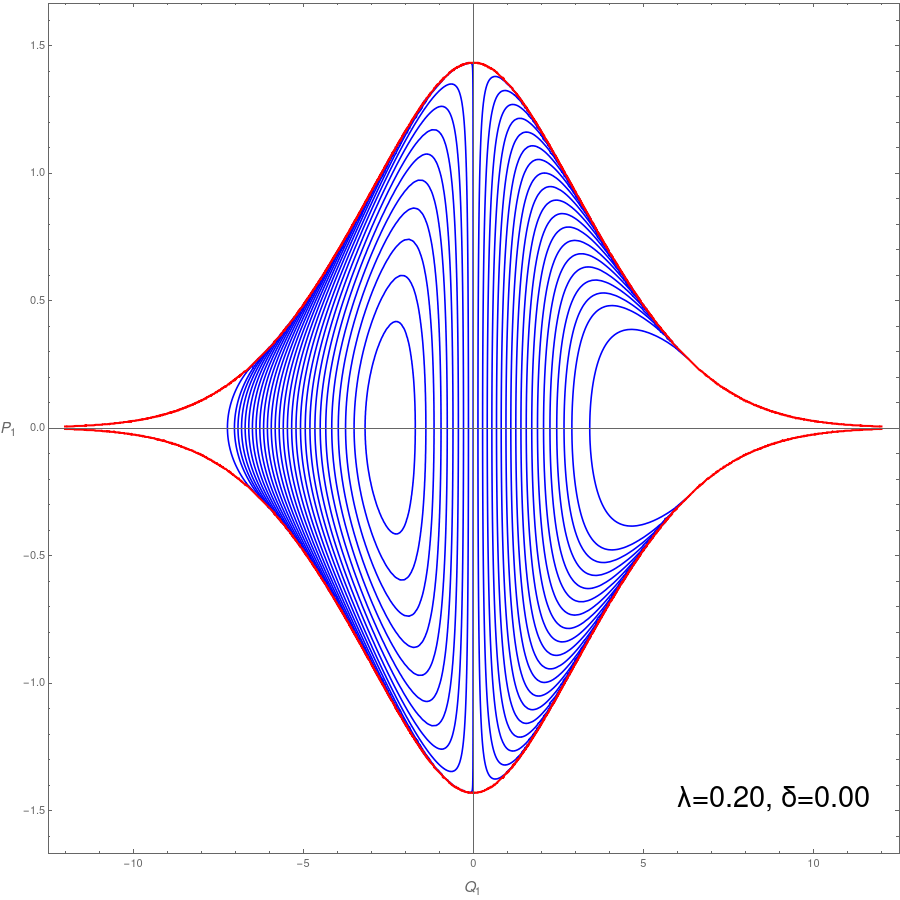}\\
\includegraphics[scale=0.135]{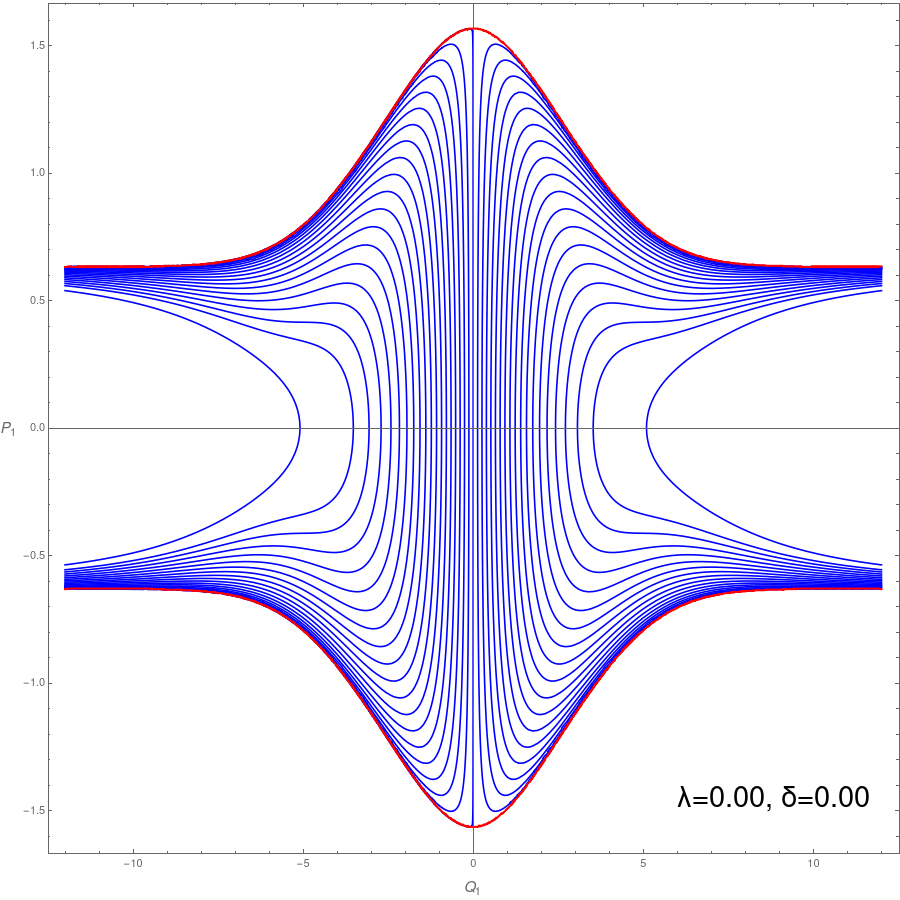}
\includegraphics[scale=0.135]{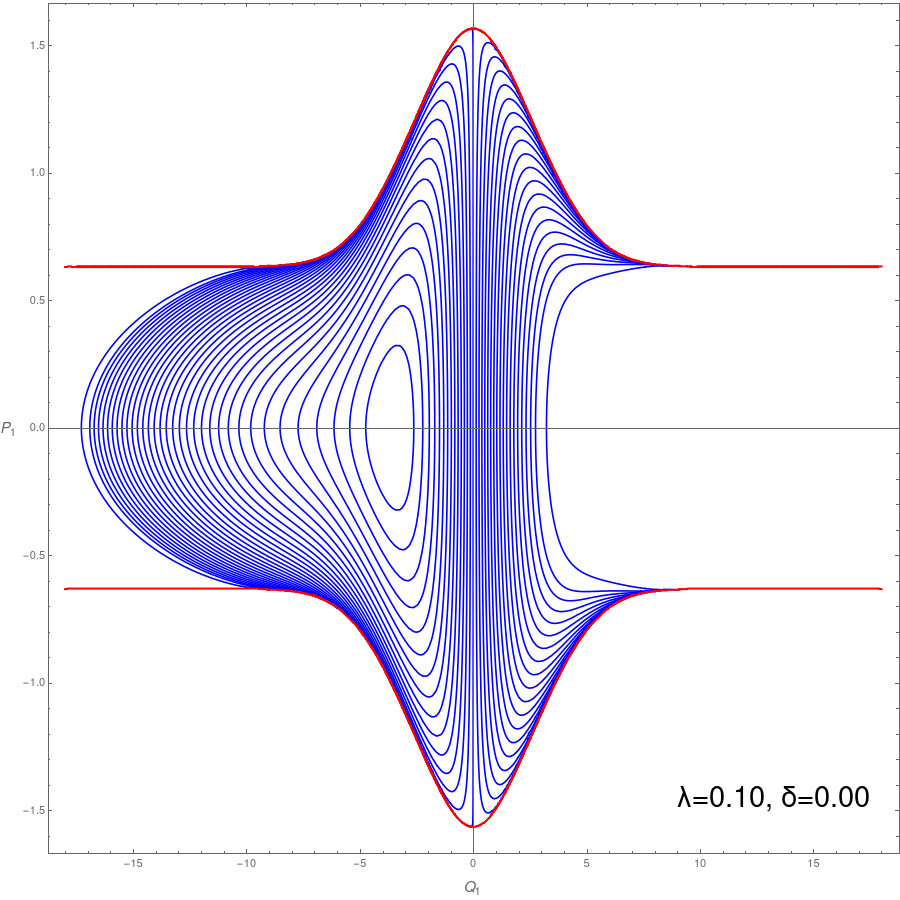}
\includegraphics[scale=0.135]{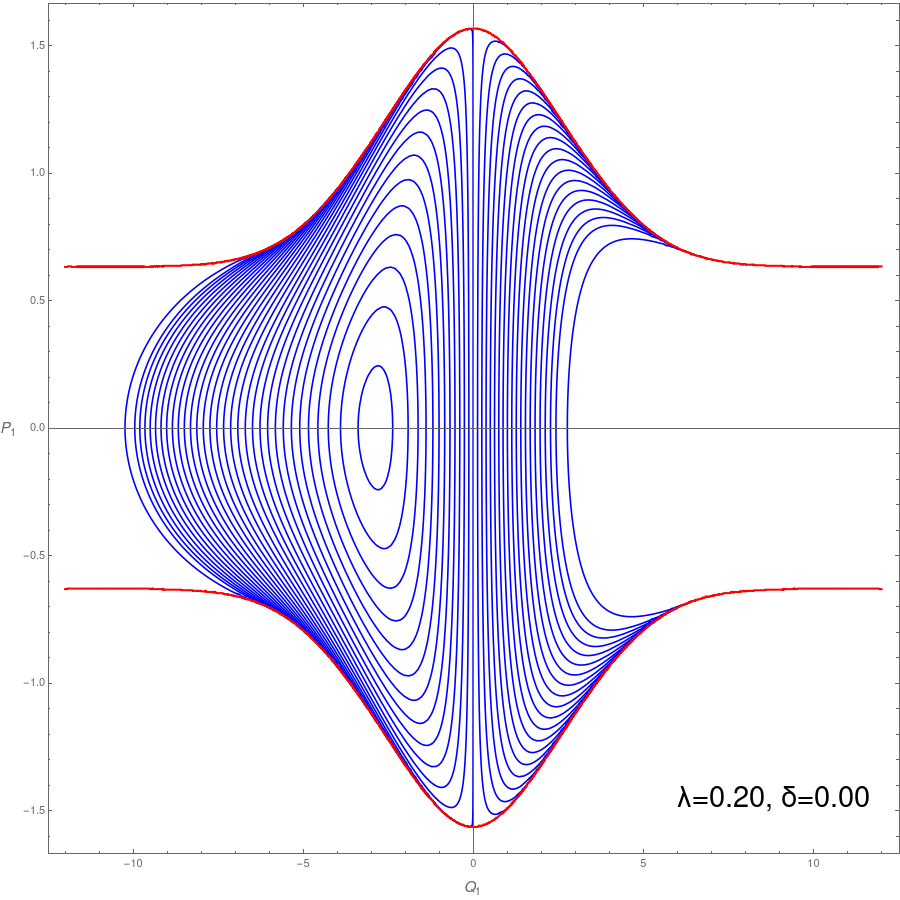}
\caption{Poincar\'e surface of section $Q_2 = 0$ for $\delta_1 =0.9$, $\delta_2=0.09$, $\delta = \delta_3-\delta_2=0$, $\kappa=1$, $E=V_{max} (top)$ and  $E>V_{max}$ (bottom) showing the influence of $\lambda$ on escape dynamics.}
\label{fig_integrable}
\end{figure}

\subsubsection{Integrable case}

We can see that for the particular choice $\delta_2 = \delta_3$, i.e.\  for isotropic distribution functions in the $p_y-p_z$ plane, the pseudopotential $V$ of  \eqref{V_sp} possesses  rotational symmetry around the origin in $(Q_1,Q_2)$. This rotational symmetry implies that the angular momentum-like function 
\begin{eqnarray}
\Lc = Q_1 P_2 - Q_2 P_1\,, \label{angular_momentum}
\end{eqnarray}
is a second integral of motion, independent of $\Hc$, and thus the canonical system \eqref{can_ham_eq_1}--\eqref{can_ham_eq_4} is integrable. Note that $\Lc$ is an integral for any potential that has the symmetry
\begin{eqnarray}
Q_1 \frac{\partial V}{\partial Q_2}=Q_2 \frac{\partial V}{\partial Q_1}\,. \label{integrability_condition}
\end{eqnarray} 
%
Let us   construct Poincar\'e surfaces of section for the case  $Q_2=0$ and crossings with $\dot{Q}_2>0$ to investigate the influence of $\lambda$ on the escape dynamics, which translates to a corresponding influence on the appearance of magnetic field fluctuations and bipolar electric field pulses. Assigning a value for the energy \eqref{Hamiltonian_2} we can  construct analytically a Poincar\'e surface of section considering various values for the angular momentum $\Lc$, e.g., on the $(Q_1,P_1)$ plane with $Q_2=0$.  The Poincar\'e surfaces of section are given in Fig. \ref{fig_integrable}. The first row corresponds to $\Hc=E_0=V_{max}$ and the second row corresponds to $E_0>V_{max}$. We observe that increasing $\lambda$ induces trapping of the phase space orbits in the latter case.

\subsubsection{Nonintegrable case and chaotic dynamics}

\begin{figure}
\centering
\includegraphics[scale=0.135]{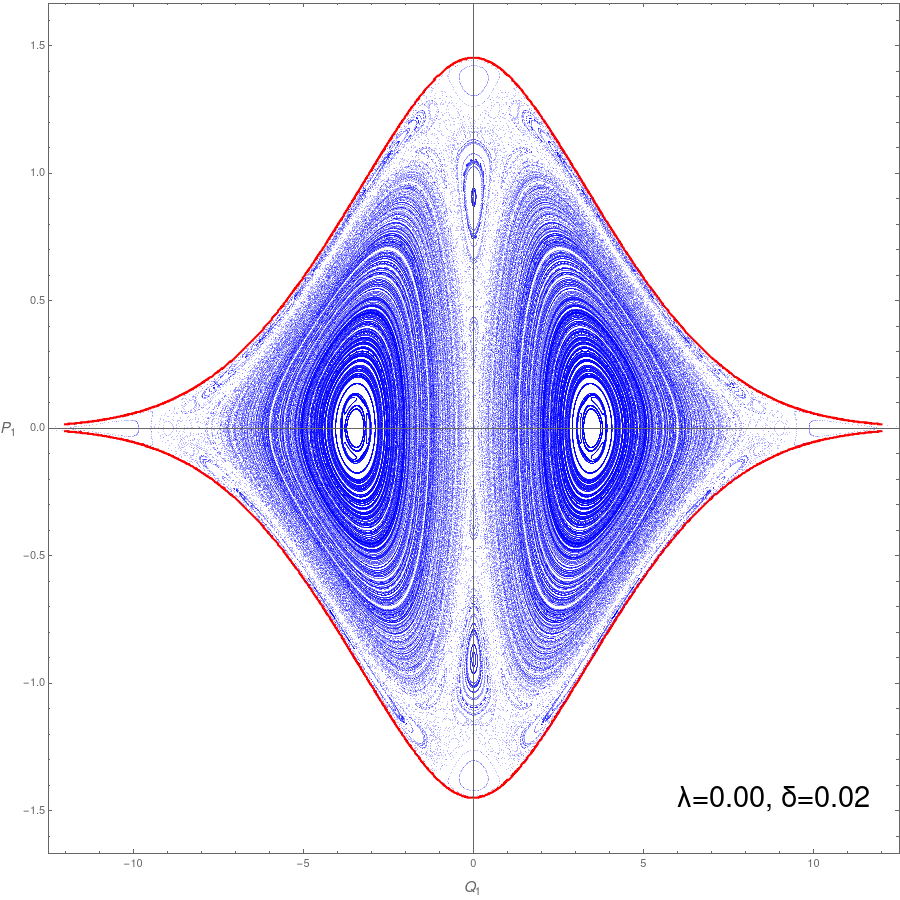}
\includegraphics[scale=0.135]{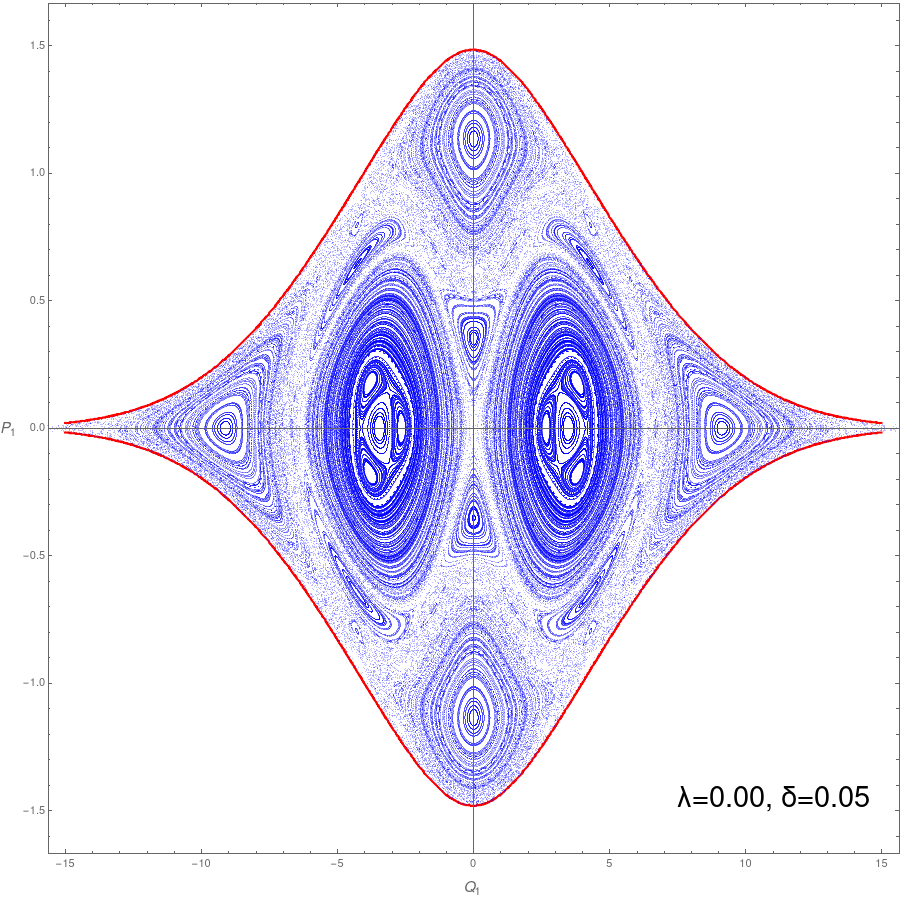} 
\includegraphics[scale=0.135]{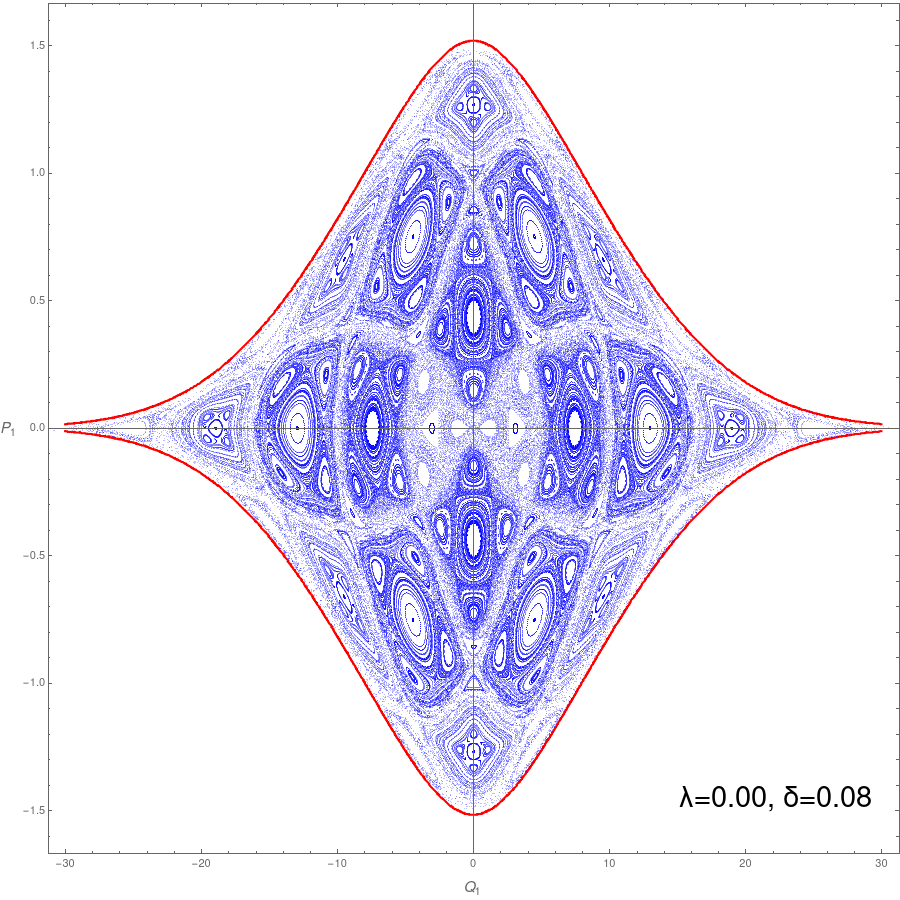} \\
\includegraphics[scale=0.135]{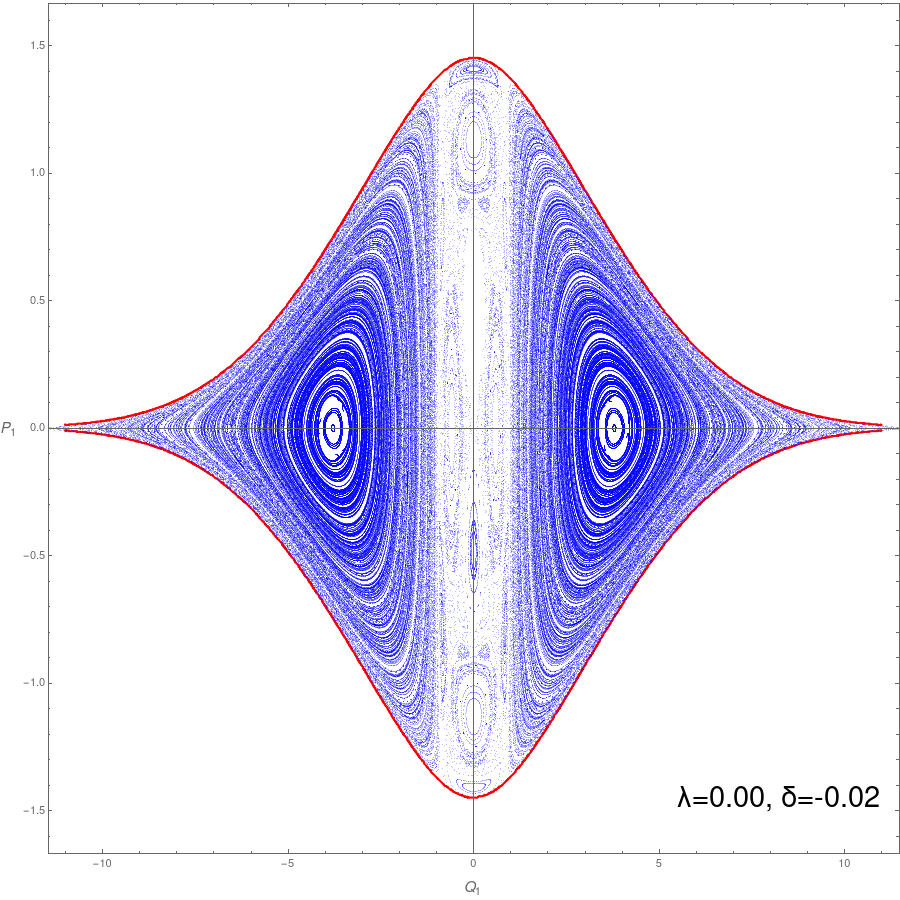}
\includegraphics[scale=0.135]{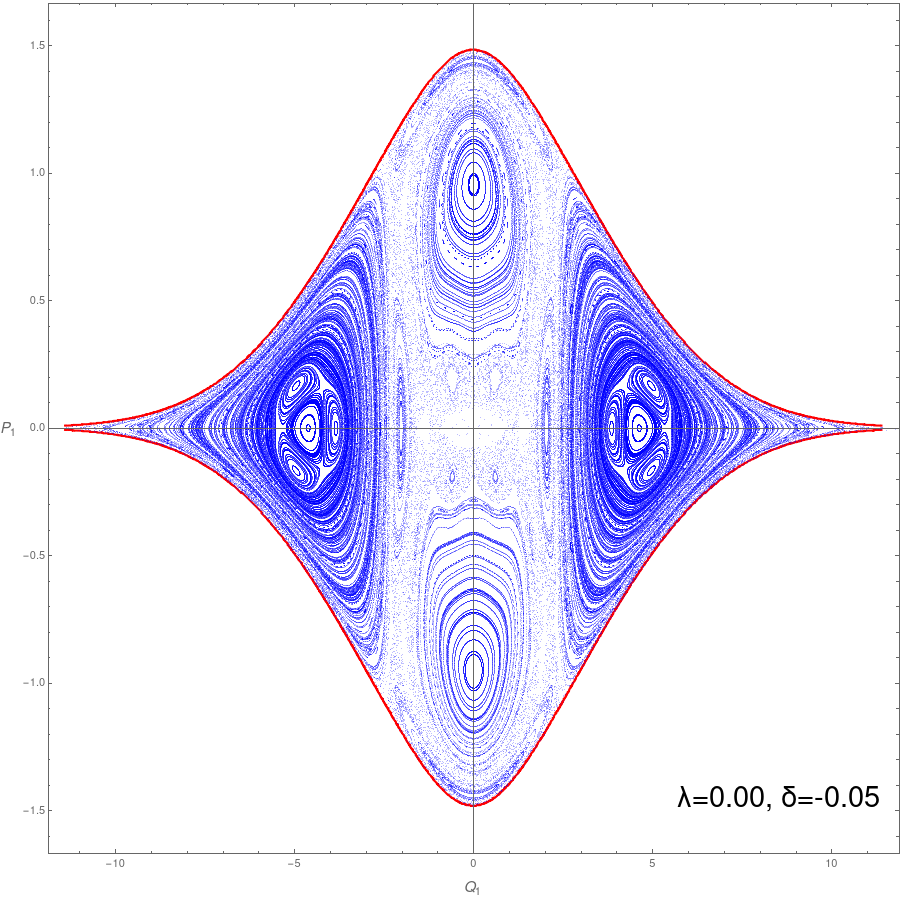}
\includegraphics[scale=0.135]{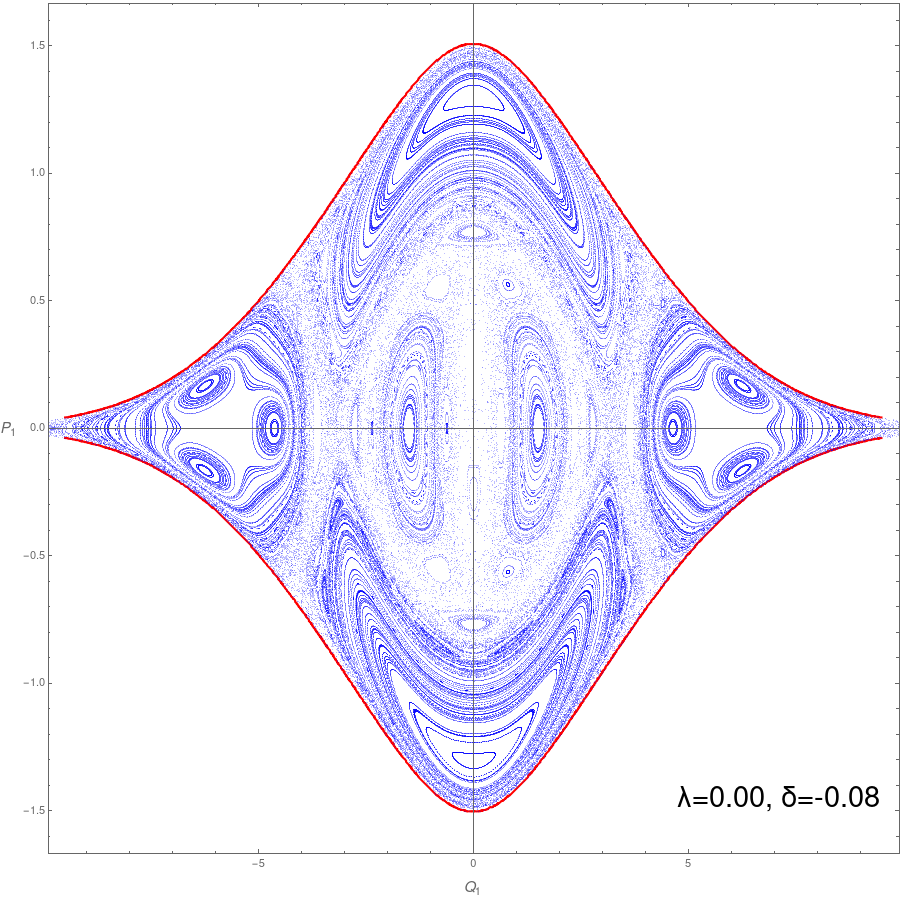}
\caption{Poincar\'e surface of section $Q_2 = 0$ for various values of the parameter $\delta = \delta_3-\delta_2\neq 0$ (nonintegrable case). The emergence of chaotic regions is observed while $|\delta|$ increases.}
\label{fig_nintegrable_D}
\end{figure}

In the generic case $\delta_2\neq \delta_3$, which corresponds to an anisotropic distribution function, the rotational symmetry of the potential is lost and the second integral of motion no longer exists. The loss of integrability results in the breakup of invariant tori and the subsequent emergence of islands of stability and chaotic regions in the Poincar\'e surfaces of section (See Fig.~\ref{fig_nintegrable_D}). As a result, the magnetic and the electric field fluctuations may be either periodic/quasiperiodic, if the initial conditions correspond to  periodic/quasiperiodic trajectories in the phase space, or chaotic if the initial conditions fall into chaotic regions. As it can be deduced from Fig.~\ref{fig_nintegrable_D}, the breakup of the invariant tori and the extent of the chaotic region depends on the  distribution function anisotropy in the $p_y-p_z$ plane, which is quantified by the parameter $\delta=\delta_3-\delta_2$. Increasing the absolute value of $\delta$ results in less organized phase-space configurations, with a large number of islands and extended chaotic regions.

\begin{figure}
\centering
\includegraphics[scale=0.135]{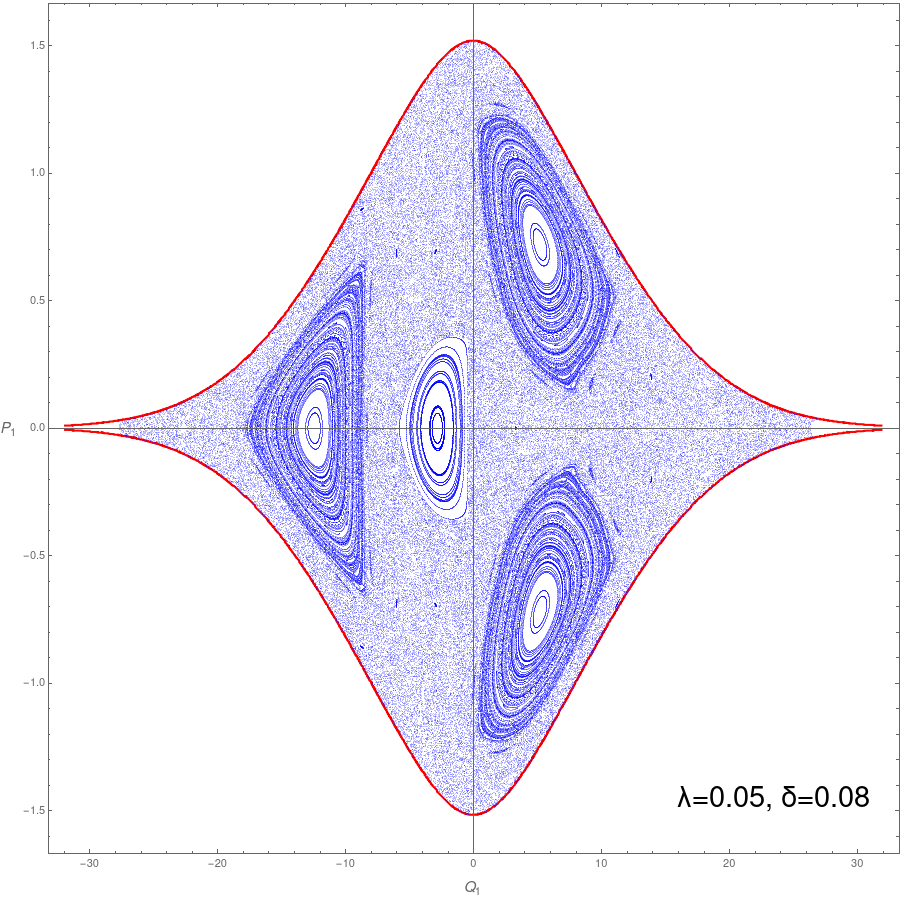}
\includegraphics[scale=0.135]{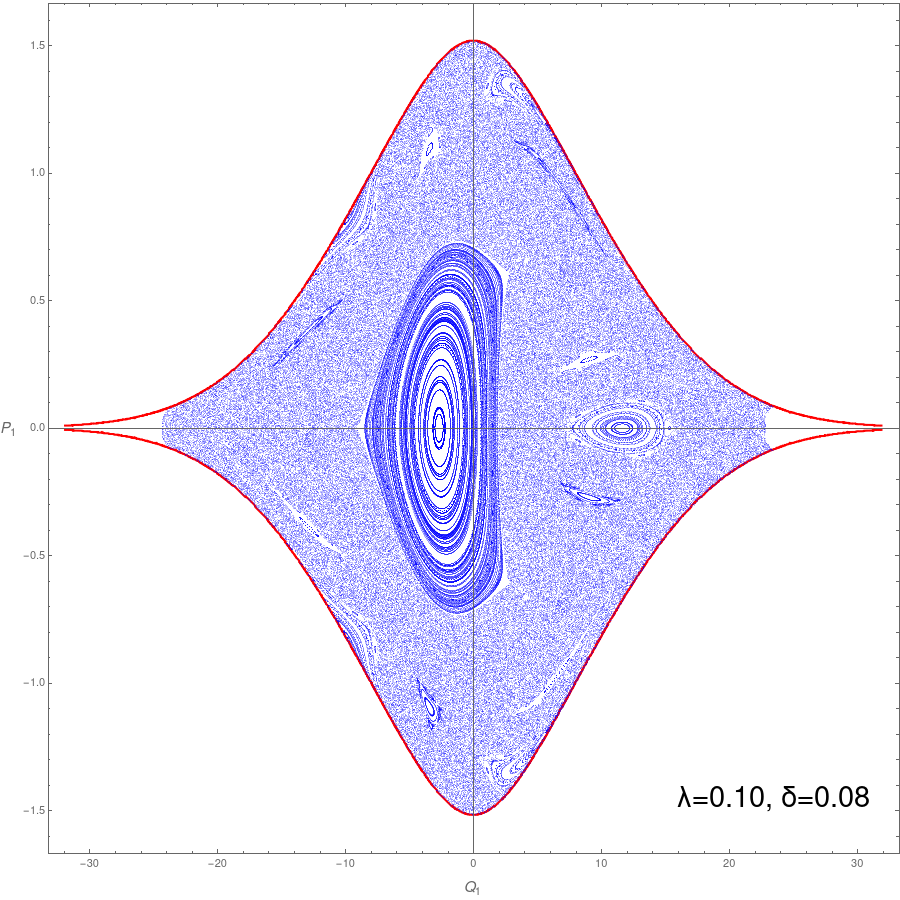}
\includegraphics[scale=0.135]{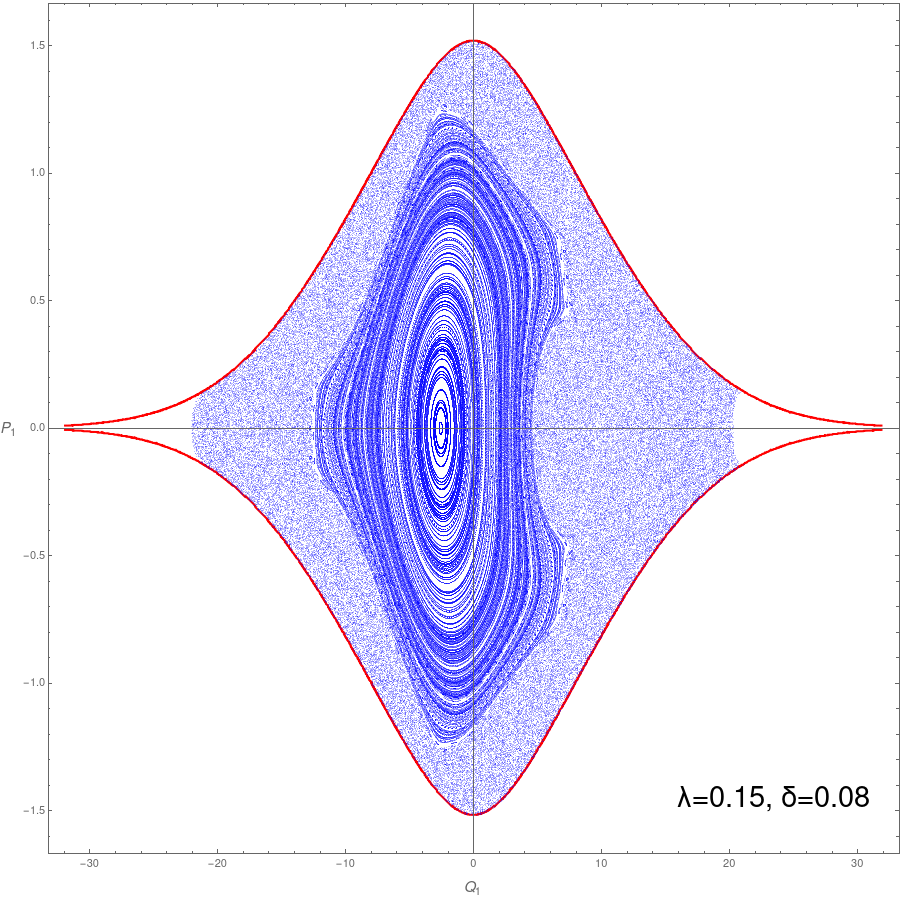}\\
\includegraphics[scale=0.135]{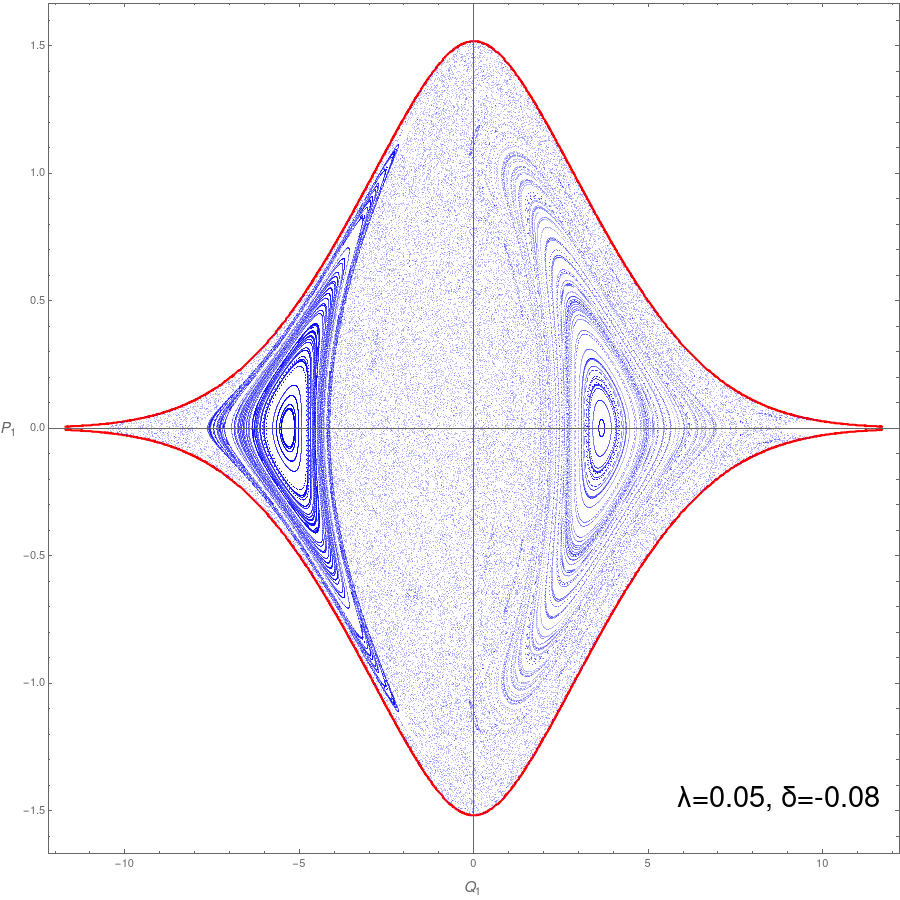}
\includegraphics[scale=0.135]{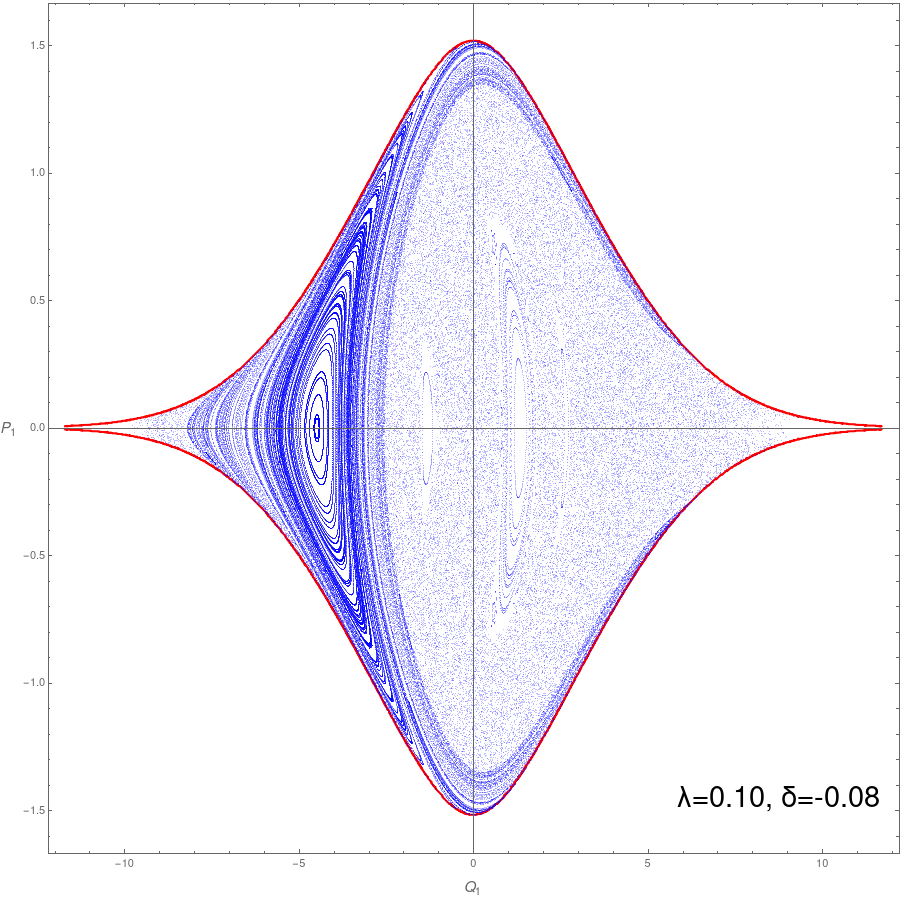}
\includegraphics[scale=0.135]{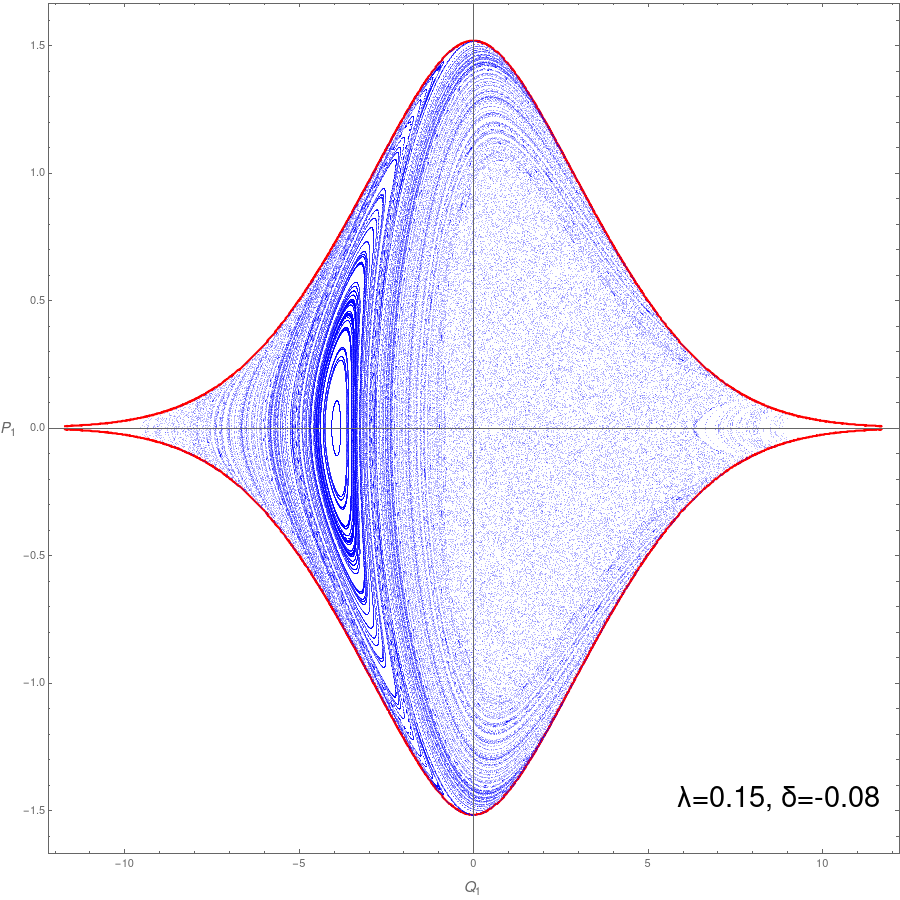}
\caption{Poincar\'e surface of section $Q_2 = 0$ of the nonintegrable system for various values of the parameter $\lambda$ and $E_0=V_{max}$.}
\label{fig_nintegrable_lambda}
\end{figure}

Regarding  the qualitative influence of the Coriolis coupling parameter (Beltrami parameter), we notice a drastic reduction in the number of islands as $\lambda$ increases (Fig.~\ref{fig_nintegrable_lambda}). More specifically as $\lambda$ increases from $0.00$ to $0.15$ the phase-space trajectories are  progressively organized into fewer, larger islands surrounded by a chaotic region, up to a point where a single extended island is formed. For the case $\delta>0$ this island covers an extended region of the Poincar\'e surface of section that contains the origin $(Q_1,P_1)=(0,0)$; thus for sufficiently small initial conditions the trajectories are stable and regular. Therefore, it can be inferred that the current density due to electrons can induce a regularization of the magnetic field dynamics and consequently a regularization of the bipolar electric field structures.

\begin{figure}
\centering
\includegraphics[scale=0.135]{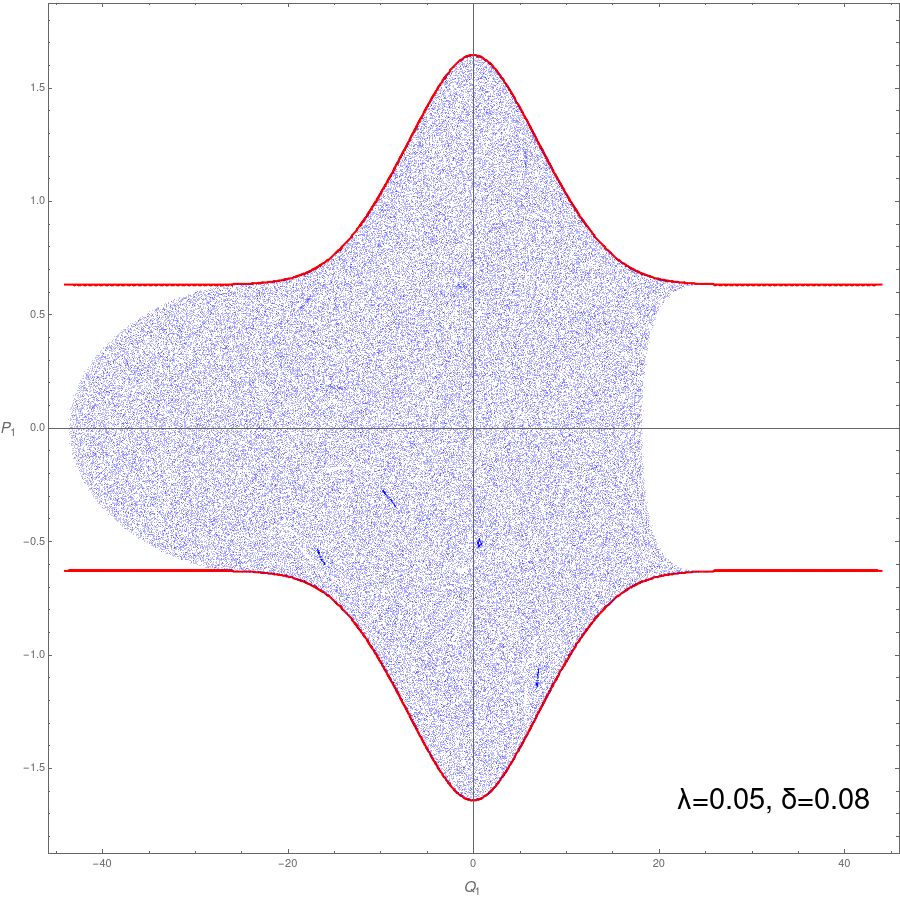}
\includegraphics[scale=0.135]{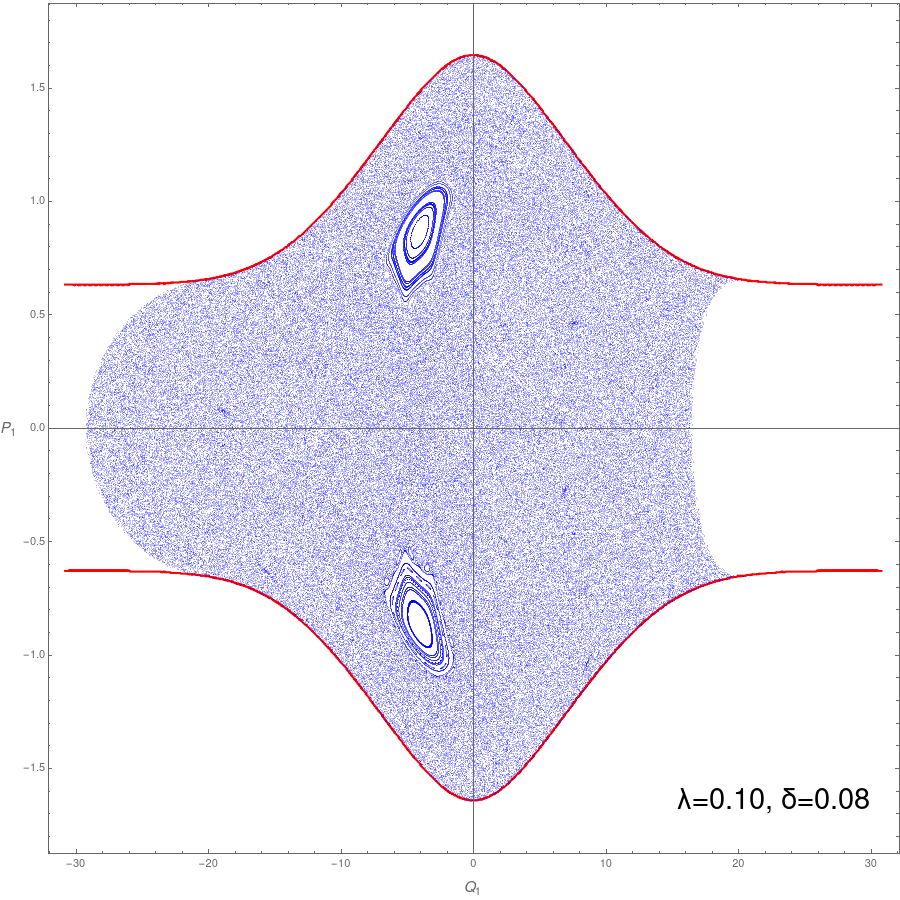}
\includegraphics[scale=0.135]{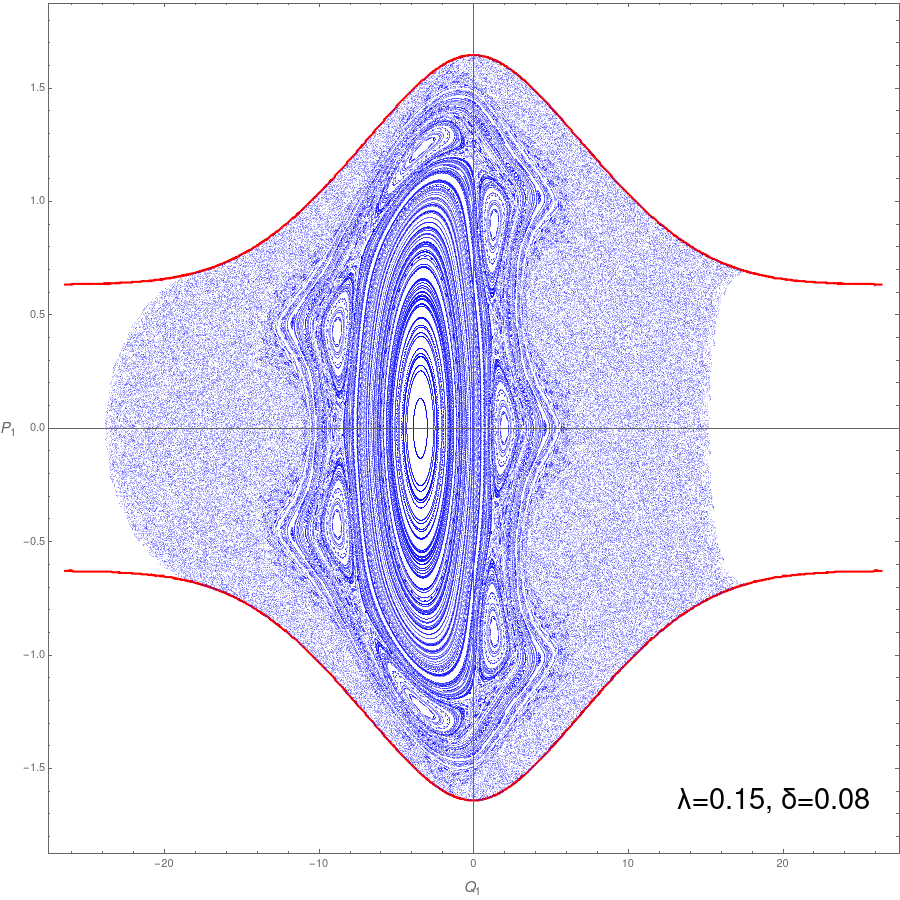}\\
\includegraphics[scale=0.135]{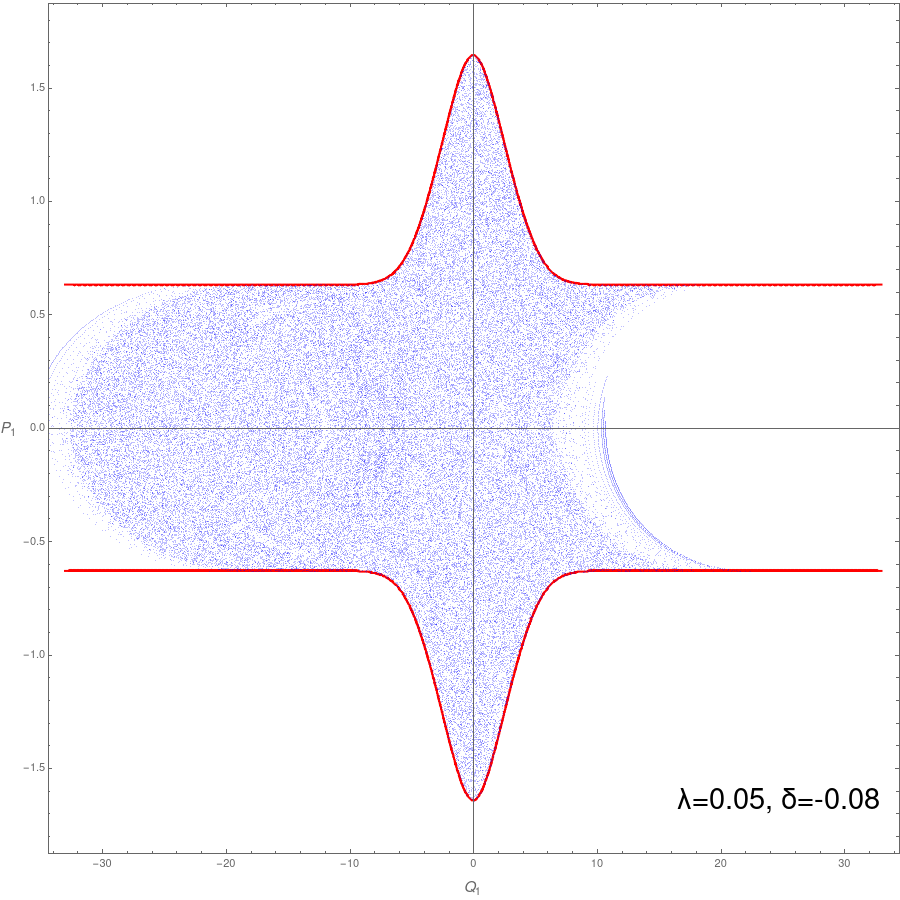}
\includegraphics[scale=0.135]{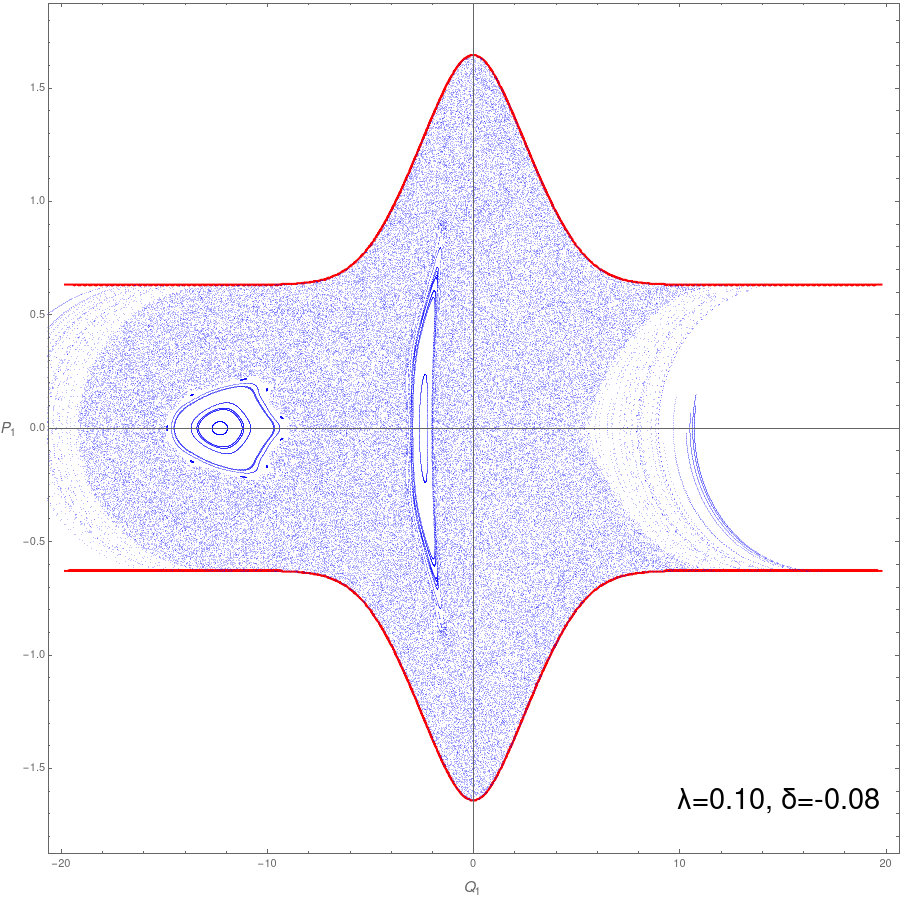}
\includegraphics[scale=0.135]{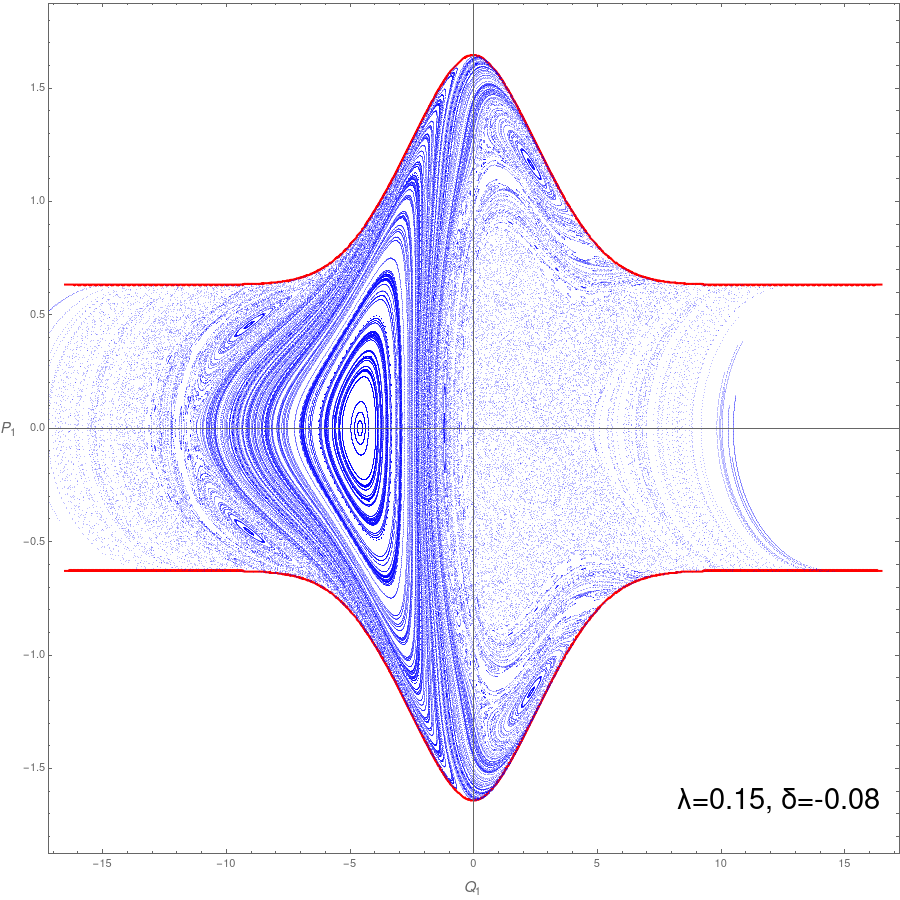}
\caption{Poincar\'e surface of section $Q_2 = 0$ of the nonintegrable system for various values of the parameter $\lambda$ and $E_0>V_{max}$.}
\label{fig_esc_nintegrable_lambda}
\end{figure}

In terms of the escape dynamics, we observe  that $\lambda$ plays a significant role in the confinement of the phase space trajectories when $E_0>V_{max}$, as shown in Figure~\ref{fig_esc_nintegrable_lambda}. This was also seen in the case in the integrable system. As $\lambda$ is increased from $0.05$ to $0.15$, phase space trajectories starting from a specific region close to the origin $(Q_1,P_1)=(0,0)$ on $Q_2=0$, occupy a progressively smaller region of the phase space and islands of stability are formed. For instance, when $\lambda=0.15$, an extended island that encloses the origin and is surrounded by a chain of smaller ones is formed, meaning that orbits originating from the vicinity of the origin will remain confined and regular. Similar findings have been reported for a rotating H\'enon-Heiles system \cite{Henon1964}, which has been studied in \cite{Salas2022}. This represents another stabilization effect of $\lambda$, with the first one having been discussed in the previous subsection, which was about the change of the stability of the equilibrium point in the case of positive electrostatic potential, i.e., $\delta_1<0$.

\begin{figure}
\centering
\includegraphics[scale=0.3]{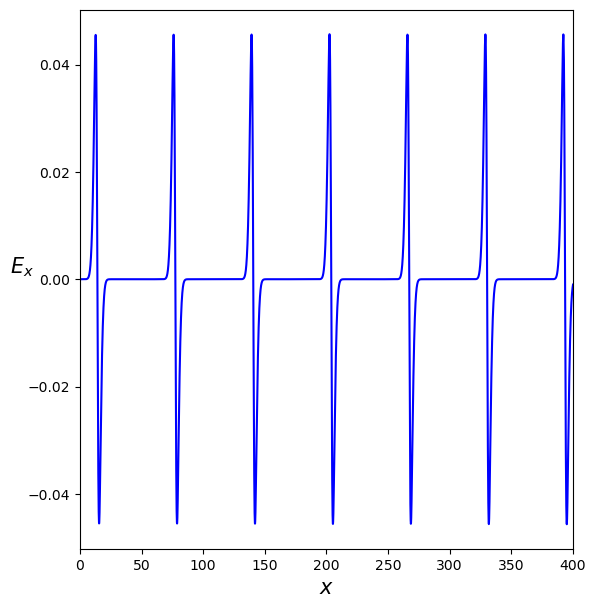}
\includegraphics[scale=0.3]{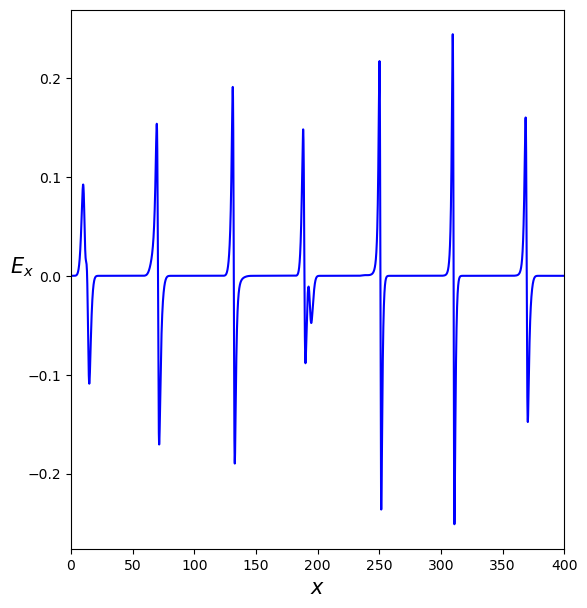}
\caption{Bipolar electric field pulses with negative electrostatic potential in the integrable (left) and the nonintegrable case (right). In the former case the pulses are periodic, while in the latter, they are irregular since the initial conditions correspond to a chaotic trajectory of the Hamiltonian system \eqref{can_ham_eq_1}--\eqref{can_ham_eq_4}. }
\label{fig_n-integrable_E}
\end{figure}

Closing this section we present two characteristic equilibria with multiple bipolar field structures in Fig. \ref{fig_n-integrable_E}. In the first case the distribution function is gyrotropic and thus the magnetic field is integrable resulting in periodic  pulses. The second corresponds to an asymmetric distribution function  ($\delta_2\neq \delta_3$), and nonintegrable magnetic field resulting in pulses with irregular spatial distribution and amplitudes. In both cases $E_0=1.05\,V_{max}$ and $\lambda=0.1$ so there is Coriolis trapping of the escaping trajectories. It has been verified that those structures emerge in positions associated with steep magnetic field gradients. Furthermore, the characteristic length scale of the pulses is determined by the distribution function parameters $\delta_1$, $\delta_2$ and $\delta_3$ and depend also on the initial conditions. The smallest structures we were able to construct have characteristic length scales of the order $0.1 \ell_i$.

\section{Harmonic oscillator potential: an  inverse equilibrium problem}
\label{sec_VI}
In the previous section we implemented the direct equilibrium construction where the pseudopotential is determined upon defining the ion distribution function. However, \eqref{V_gen} can be seen as an integral equation for determining the function $g(p_y,p_z)$ in the expression for the distribution function, for a known pseudopotential $V(Q_1.Q_2)$. For example,  we may choose a pseudopotential that enables the calculation of analytic solutions for the scalar and vector potentials and then solve the integral equation \eqref{V_gen} for $g(p_y,p_z)$. In order to obtain a wave-like solution let us adopt a two-dimensional (2-D) anisotropic harmonic oscillator potential of the form
\begin{eqnarray}
    V(Q_1,Q_2)= V_0+ \frac{\omega_1^2}{2} Q_1^2+\frac{\omega_2^2}{2} Q_2^2\,, \label{V_harm_osc}
\end{eqnarray}
where $V_0$, $\omega_1$ and $\omega_2$ are real constants. In this case the canonical Hamiltonian system \eqref{can_ham_eq_1}--\eqref{can_ham_eq_4} can be written in terms of the generalized coordinates $Q_1$, $Q_2$ as:
\begin{eqnarray}
    \Ddot{Q}_1 -\lambda \dot{Q}_2 +\omega_1^2 Q_1 =0\,,\label{ddotQ1}\\
    \Ddot{Q}_2+\lambda \dot{Q}_1 +\omega_2^2 Q_2=0\,. \label{ddotQ2}
\end{eqnarray}
The system of   \eqref{ddotQ1}--\eqref{ddotQ2}  possesses the following analytic solution
\begin{eqnarray}
    Q_1(t) &=& \sum_{i=1}^4 c_i e^{\eta_it}\,,\label{an_sol_Q1}\\
    Q_2(t) &=& \sum_{i=1}^4 \frac{\eta_i^2 + \omega_1^2}{\lambda \eta_i}c_{i} e^{\eta_it}\,, \label{an_sol_Q2}
\end{eqnarray}
where $\eta_{1,2,3,4}$ are given in Appendix \ref{app_1} and  $c_i$, $i=1,...,4$ are arbitrary constants that are determined by the initial conditions imposed on the canonical variables  $(Q_1,Q_2,P_1,P_2)$. The canonical momenta $P_1$, $P_2$ can be easily derived from the definitions \eqref{canonical_vars} and \eqref{an_sol_Q1}, \eqref{an_sol_Q2}.

Therefore, as is elementary,  the Hamiltonian system with a 2-D anisotropic harmonic oscillator potential is integrable and solvable,  although the angular momentum $\Lc$ is not conserved for $\omega_1\neq \omega_2$. Obviously, the Hamiltonian $\Hc$ remains an integral of motion. If the oscillators were uncoupled ($\lambda =0$), the Hamiltonian would be diagonal, and the energy stored in one of the degrees of freedom, $\Hc_1 = P_1^2/2+\omega_1^2 Q_1^2/2$, would be a second integral of motion. However, for $\lambda\neq 0$, the Hamiltonian is seemingly non-separable, and the second integral of motion is  not directly apparent. Nonetheless, because the system is linear it must be  transformable into one of the normal forms for linear systems  \cite{moser58} by a linear canonical transformation $(Q_1,Q_2,P_1,P_2)\rightarrow (\tilde{Q}_1,\tilde{Q}_2,\tilde{P}_1,\tilde{P}_2)$.  If the system is stable, this amounts to a transformation to action-angle variables where the Hamiltonian is the sum of uncoupled oscillators. This separable form of the Hamiltonian reveals the second integral of motion. Specifically, by employing the canonical transformation presented in Appendix \ref{app_1} (Eqs. \eqref{generating_fun}--\eqref{canonical_transf_2}), we can show that in the stable case, the Hamiltonian takes the form  
\begin{eqnarray}
    \tilde{\Hc} - V_0 =  \frac{1}{2}\sum_{i=1,2} \sigma_i \zeta_i (\tilde{Q}_i^2 + \tilde{P}_i^2)\,.\label{normal_form_ham}
\end{eqnarray}
where, $\sigma_i\in\{\pm1\}$  denotes the  signature of the modes with frequencies $\zeta_i$. In Appendix \ref{app_1} we show that $\sigma_1=-1$, thus $\zeta_1=\zeta_-$ corresponds to a negative energy mode. Due to the large parameter space, we will not explore this further here and instead focus on the inverse equilibrium problem.

With the pseudopotential \eqref{V_harm_osc} the  integral equation \eqref{V_gen} reads
\begin{eqnarray}
\int_{-\infty}^{\infty} dp_y \int_{-\infty}^{\infty} dp_z e^{-\frac{1}{2}(p_y-Q_1)^2-\frac{1}{2}(p_z-Q_2)^2}g(p_y,p_z) \nn\\
= \frac{\sqrt{2}}{\sqrt{\pi}(\kappa+1)^{\kappa+1}}\left(V_0+\frac{\omega_1^2}{2}Q_1^2 + \frac{\omega_2^2}{2} Q_2^2\right)^{\kappa+1}\,. \label{integral_eq_1}
\end{eqnarray}
Whenever the right hand side of \eqref{integral_eq_1} can be written as a polynomial of $Q_1$ and $Q_2$ then there is a quite expedient method to compute the kernel $g(p_y,p_z)$ and consequently the distribution function $f(H,p_y,p_z)$, using Hermite polynomials (see \cite{Channell1976,Allanson2016}). The rhs of \eqref{integral_eq_1} can be expressed as a polynomial either if $\kappa$ is a positive integer (or zero in the cold electron limit) or  if $\omega_1^2 Q_1^2/2 + \omega_2^2 Q_2^2/2 < V_0$ since in that case its Taylor series is convergent.  In both cases, using the following relation \cite{Channell1976}:
\begin{eqnarray}
    e^{-\frac{1}{2}(x-y)^2} = \sum_{n} \frac{e^{-x^2/2}}{n!} H_n\left( \frac{x}{\sqrt{2}}\right) \left( \frac{y}{\sqrt{2}}\right)^n\,,
\end{eqnarray}
where $H_{n}(x)$ is the $n$th-order Hermite polynomial defined by
\begin{eqnarray}
    H_n(x)=(-1)^n e^{x^2} \frac{d^n}{dx^n} e^{-x^2/2}\,.
\end{eqnarray}
Equation \eqref{integral_eq_1} can be expanded as follows:
\begin{eqnarray}
\sum_{m,n} \frac{Q_1^n Q_2^m}{2^{m+n} m! n! } \int_{-\infty}^{\infty} dp_y \int_{-\infty}^{\infty} dp_z e^{-\frac{p_y^2}{2}-\frac{p_z^2}{2}}H_n\left(\frac{p_y}{\sqrt{2}}\right)H_m\left(\frac{p_z}{\sqrt{2}}\right) g(p_y,p_z) \nn \\
= \sqrt{\frac{2}{\pi}} \left(\frac{V_0}{\kappa+1}\right)^{\kappa+1} \sum_{\mu} \sum_{\nu=0}^{\mu} \begin{pmatrix}
    \kappa+1 \\ \mu
\end{pmatrix} \begin{pmatrix}
    \mu \\   \nu
\end{pmatrix}
\frac{\omega_1^{2(\mu-\nu)}\omega_2^{2\nu}}{(2V_0)^\mu} Q_1^{2(\mu-\nu)}Q_2^{2\nu}\,. \label{binomial}
\end{eqnarray}
 Since the Hermite polynomials satisfy the following orthogonality conditions: 
\begin{eqnarray}
    \int_{-\infty}^{\infty} dx H_m(x) H_n(x) e^{-x^2} = \sqrt{\pi}2^n n! \delta_{mn}\,, \label{hermite_orthogonality}
\end{eqnarray}
it is appropriate to expand the kernel $g(p_y,p_z)$ using a Hermite polynomial basis, i.e.,
\begin{eqnarray}
    g(p_y,p_z) = \sum_{k,\ell} c_{k\ell}H_k\left(\frac{p_y}{\sqrt{2}}\right) H_\ell\left(\frac{p_z}{\sqrt{2}}\right)\,. \label{g_hermite}
\end{eqnarray}
Inserting \eqref{g_hermite} into \eqref{binomial} and using the orthogonality of Hermite polynomials \eqref{hermite_orthogonality} we are left with
\begin{align}
&  \sum_{m,n} c_{nm} Q_1^n Q_2^m =\frac{\sqrt{2}}{\pi^{3/2}} \left(\frac{ V_0}{\kappa +1}\right)^{\kappa+1} \times \nn \\
 & \times \sum_{\mu}\sum_{\nu=0}^{\mu}\begin{pmatrix}
        \kappa+1 \\ \mu
    \end{pmatrix} \begin{pmatrix}
        \mu \\ \nu
    \end{pmatrix}
    \frac{\omega_1^{2(\mu-\nu)}\omega_2^{2\nu}}{(2V_0)^\mu}Q_1^{2(\mu-\nu)}Q_2^{2\nu}\,; 
\end{align}
thus, the only nonvaninshing coefficients $c_{nm}$ in the expansion \eqref{g_hermite} are
\begin{equation}
   c_{2(\mu-\nu),2\nu}= 
        \frac{\sqrt{2}}{\pi^{3/2}}\left( \frac{V_0}{\kappa+1}\right)^{\kappa+1} 
        \begin{pmatrix}
        \kappa+1 \\ \mu
    \end{pmatrix} 
    \begin{pmatrix}
     \mu \\ \nu
    \end{pmatrix}
    \frac{\omega_1^{2(\mu-\nu)}\omega_2^{2\nu}}{(2V_0)^\mu}\,, \quad
    \nu =0,...,\mu\,.
\end{equation}
In addition, we have shown that  (see Eqs.~\eqref{Phi_qn}, \eqref{G_fun} and \eqref{V_gen})  the electrostatic potential is given by 
\begin{eqnarray}
 \Phi(Q_1,Q_2) = ln \left(\frac{V(Q_1,Q_2)}{\kappa+1}\right)^{\kappa}\,; 
\end{eqnarray}
thus,  for the anisotropic harmonic oscillator pseudopotential \eqref{V_harm_osc}, $\Phi$ reads as follows:
\begin{eqnarray}
    \Phi(Q_1,Q_2) =   ln \left(\frac{V_0}{\kappa+1}+ \frac{\omega_1^2}{2(\kappa+1)}Q_1^2 + \frac{\omega_2^2}{2(\kappa+1)}Q_2^2 \right)^\kappa\,.
\end{eqnarray}
Consequently, the equilibrium distribution function is
\begin{align}
  &  f(x,v_x,v_y,v_z) = exp\left(-\frac{v^2}{2} - \Phi \right) g(p_y,p_z) \nn\\
   & = \frac{\sqrt{2} }{\pi^{3/2}(\kappa+1)} e^{-v^2/2}\frac{V_0^{\kappa+1}}{\left(V_0+\frac{\omega_1^2}{2}Q_1^2+\frac{\omega_2^2}{2} Q_2^2\right)^{k}}\times \nn \\ 
   & \times \sum_\mu \sum_{\nu=0}^{\mu} \begin{pmatrix}
        \kappa+1 \\ \mu 
    \end{pmatrix} \begin{pmatrix}
        \mu \\ \nu
    \end{pmatrix} \frac{\omega_1^{2(\mu-\nu)}\omega_2^{2\nu}}{(2 V_0)^\mu} H_{2(\mu-\nu)}\left(\frac{p_y}{\sqrt{2}} \right)H_{2\nu}\left(\frac{p_z}{\sqrt{2}} \right)\,.
\end{align}
As  examples, consider two special cases:  first, the cold electron limit, i.e. $\kappa=0$,  which implies $\Phi=0$ (strong neutrality); and second, the case $\kappa=1$. In the former case, which has been considered also in \cite{Channell1976} (for the isotropic oscillator potential) the equilibrium distribution function reads as
\begin{eqnarray}
f(x,\bsv) &=& \frac{\sqrt{2}}{\pi^{3/2}}e^{-v^2/2}\Big[V_0 - (\omega_1^2+\omega_2^2)
\nn\\
&& \hspace{1cm} + \omega_1^2 (v_y+Q_1)^2 + \omega_2^2 (v_z+Q_2)^2\bigg]\,.
\end{eqnarray}
The positivity of the distribution function is ensured if $V_0>\omega_1^2+\omega_2^2$. In the latter case ($\kappa=1$) the equilibrium distribution function is given by
\begin{eqnarray}
    f(x,\bsv)&=& \frac{\sqrt{2}}{2\pi^{3/2}}\frac{e^{-v^2/2}}{V_0 + \frac{\omega_1^2}{2}Q_1^2+\frac{\omega_2^2}{2}Q_2^2}\Big[c_0 + c_1 (v_y+Q_1)^2 + c_2 (v_z+Q_2)^2\nn\\
   &+& c_3 (v_y+Q_1)^2(v_z+Q_2)^2 + c_4 (v_y+Q_1)^4+ c_5 (v_z+Q_2)^4\Big]\,,
\end{eqnarray}
where
\begin{eqnarray}
    c_0 = V_0^2 - 2 V_0 \omega_1^2- 2V_0 \omega_2^2+ 2 \omega_1^2 \omega_2^2 +3 \omega_1^4 + 3 \omega_2^4\,,\nn\\
    c_1 = 2V_0 \omega_1^2 - 2 \omega_1^2 \omega_2^2 - 6 \omega_1^4\,,\nn\\
    c_2 = 2V_0\omega_2^2 - 2 \omega_1^2 \omega_2^2 - 6 \omega_2^2\,,\nn\\
    c_3= 2\omega_1^2 \omega_2^2\,, \quad c_4= \omega_1^4\,, \quad c_5 = \omega_2^4\,.
\end{eqnarray}
One can find regions in the parameter space $(V_0,\omega_1,\omega_2)$, in which the distribution function is positively defined.

\section{Conclusion}
\label{sec_VII}

We calculated equilibrium solutions to the hybrid Vlasov-Maxwell model with kinetic ions and massless fluid electrons by solving an inhomogeneous Beltrami equation, where the inhomogeneous term is the kinetic ion current density. To this end, we used the quasineutrality condition to express the electrostatic potential in terms of the vector potential components,  instead of assuming strong neutrality to allow the emergence of large amplitude electric field structures due to fluctuations in the magnetic field. Then, we studied the Hamiltonian dynamics of the magnetic field equations and found that fluctuations are regular for ion distribution functions characterized by rotational symmetry in the $p_y-p_z$ plane, but can become chaotic for asymmetric ones. We have shown that the magnetic field equations constitute a Hamiltonian system and we have used Poincar\'e maps to show the existence of periodic, quasiperiodic and chaotic trajectories in phase-space, indicating that magnetic and electric field fluctuations can be either ordered or chaotic, depending on the initial conditions for the magnetic field and vector potential components. The electron current density affects the dynamics of the Hamiltonian system by altering the structure of the phase-space and the escape dynamics of the orbits through a Coriolis-like coupling between the generalized coordinates and the corresponding canonical momenta. We observed that increasing electron current density progressively organizes phase space trajectories into large islands of stability and can induce the trapping of escaping phase-space orbits. The stabilizing effect of the electron current density was also ascertained upon performing  a stability analysis of the equilibrium point. Upon increasing $\lambda$,  the orbits in a neighborhood of the equilibrium point can be stabilized even for positive electrostatic potentials. In this case, the electrostatic potential is modulated by the magnetic field fluctuations, enabling ion trapping. Finally, we provided an inverse equilibrium calculation where we computed the vector potential components and the ion distribution function by defining the pseudopotential of the Hamiltonian magnetic field dynamics. The particular choice of pseudopotential function results in a rotating, 2-D, harmonic oscillator Hamiltonian which is integrable and solvable. The distribution function is constructed assuming a multiplicative separability in $p_y$ and $p_z$ and using a Hermite expansion method. In future work, we plan to investigate the properties of Hamiltonian systems emerging in kinetic equilibrium problems, including the study of translationally symmetric plasmas, deviations from quasineutrality, and the existence of multiple ion species with fluid and kinetic components (e.g. \cite{Kaltsas2021}). In these frameworks the development of models capable of describing bipolar structures that closely resemble those observed by the MMS mission will be pursued.

\section*{Acknowledgements}
The authors would like to thank Anna Tenerani for providing insightful comments that helped to improve the manuscript.

\section*{Funding}
This work has received funding from the National Fusion Programme of the Hellenic Republic -- General Secretariat for Research and Innovation. P.J.M. was supported by the U.S. Department of Energy Contract No. DE-FG05-80ET-53088

\section*{Declaration of interests}
The authors report no conflict of interest.
\appendix

\section{Normal form Hamiltonian}
\label{app_1}

In section \ref{sec_IV} we found a canonical Hamiltonian that describes the magnetic field dynamics in 1-D hybrid Vlasov equilibria. Upon Taylor expanding the potential of this Hamiltonian  we obtain
\begin{eqnarray}
\bar{\Hc}= \frac{1}{2}(P_1^2+P_2^2)+\frac{\lambda^2}{8}(Q_1^2+Q_2^2)+\frac{\lambda}{2}(P_1Q_2-P_2Q_1)\nn\\
+ V_0 + \frac{V_1}{2} Q_1^2 + \frac{V_2}{2} Q_2^2 +U(Q_1,Q_2)\,,
\end{eqnarray}
where $V_0 = V(\xi_e)$ and
$$V_1=\frac{\partial^2 V}{\partial Q_1^2}(\xi_e)\,, \quad V_2=\frac{\partial^2 V}{\partial Q_2^2}(\xi_e)\,.$$ 

Hereafter the overbar will be dropped. For the pseudopotential \eqref{V_sp}, $V_1$ and $V_2$ are given by Eqs. \eqref{V_1-V_2_nonlin} and $U(Q_1,Q_2)$ represents the $\Oc(Q^3)$ terms. For the harmonic oscillator potential \eqref{V_harm_osc} in section \ref{sec_VI}, $V_1=\omega_1^2$, $V_2=\omega_2^2$ and $U=0$.

To assess the stability of the fixed point $\xi_e=(0,0,0,0)$ and in order to write the Hamiltonians in normal, diagonal form, we have to calculate the Hamiltonian spectrum of the equilibrium point. The eigenvalues of the stability matrix at $\xi_e$ are given by
\begin{eqnarray}
\eta_{j}= \pm \sqrt{\frac{A}{2} \pm \frac{\sqrt{\Delta}}{2}}= \pm i \zeta_\pm\,,\quad j=1,...,4\,.\label{eigs}
\end{eqnarray}
Here,
\begin{eqnarray}
\Delta &=& A^2 - 4B\,,\\
A&=& \frac{1}{2}Tr\left([\Jc_c (D^2 \Hc)(\xi_e)]^2\right)\,,\\
B&=& det\left(\Jc_c (D^2 \Hc)(\xi_e)\right)\,,
\end{eqnarray}
where $D^2\Hc(\xi_e)$ is the Hessian of the Hamiltonian $\Hc$ evaluated at the equilibrium point $\xi_e$:
\begin{eqnarray}
(D^2 \Hc)(\xi_e)=\left(\frac{\partial^2 \Hc}{\partial \xi_i \partial \xi_j}\right)(\xi_e)= 
\begin{pmatrix}
\frac{\lambda^2}{4}+ V_1 && 0 && 0 && -\frac{\lambda}{2}\\
0 && \frac{\lambda^2}{4}+V_2&&\frac{\lambda}{2}&&0
\\
0 && \frac{\lambda}{2}&& 1 && 0\\ 
-\frac{\lambda}{2}&& 0 && 0 && 1
\end{pmatrix}\,,
\end{eqnarray}
The eigenvalues \eqref{eigs} are thus given by 
\begin{eqnarray}
    \eta_{j} = \pm \frac{1}{\sqrt{2}} \sqrt{-\lambda^2-V_1-V_2 \pm \sqrt{(\lambda^2+V_1+V_2)^2-4V_1V_2}}\,,
\end{eqnarray}
and the two frequencies are
$$\zeta_\pm = \frac{1}{\sqrt{2}} \sqrt{\lambda^2+V_1+V_2 \pm \sqrt{(\lambda^2+V_1+V_2)^2-4V_1V_2}}\,.$$

Following \cite{pjmK90}, in the stable case ($\zeta_\pm \in \mathbb{R}$), we can diagonalize the Hamiltonian through a linear canonical transformation introducing the following generating function of type-2
\begin{eqnarray}
    F_2(Q_1,Q_2,\tilde{P}_1,\tilde{P}_2)= \alpha_1 Q_1\tilde{P}_1 + \alpha_2 Q_2\tilde{P}_2 + \alpha_3 Q_1Q_2 + \alpha_4 \tilde{P}_1 \tilde{P}_2\,,\label{generating_fun}
\end{eqnarray}
with
\begin{eqnarray}
    \alpha_1 &=& \left[ \frac{\Delta \left((V_1-V_2)^2+\lambda^2(V_1+V_2)+(V_1-V_2)\sqrt{\Delta}\right)}{2V_2\lambda^4}\right]^{1/4}\,,\nn\\
    \alpha_2 &=& \left[\frac{2 V_2 \Delta}{(V_1-V_2)^2+\lambda^2(V_1+V_2)-(V_1-V_2)\sqrt{\Delta}}\right]^{1/4}\,,\nn \\
    \alpha_3 &=& \frac{V_1-V_2 +\sqrt{\Delta}}{2\lambda}\,,\nn\\
    \alpha_4 &=&\left[\frac{(V_1-V_2)^2+\lambda^2(V_1+V_2)+(V_1-V_2)\sqrt{\Delta}}{(V_1-V_2)^2+\lambda^2(V_1+V_2)-(V_1-V_2)\sqrt{\Delta}}\right]^{1/4} \,. 
\end{eqnarray}
In the isotropic case $V_1=V_2$ we have 
\begin{eqnarray}
    \alpha_1 =\alpha_2=\left(\frac{\Delta }{\lambda^2}\right)^{1/4}\,, \quad
    \alpha_3 =\frac{\sqrt{\Delta}}{2\lambda}\,, \quad \alpha_4 = 1\,. 
\end{eqnarray}
The new coordinates and momenta can be computed by the following equations
\begin{eqnarray}
    \tilde{Q}_i = \frac{\partial F_2}{\partial \tilde{P_i}}\,, \quad
    P_i = \frac{\partial F_2}{\partial Q_i}\,, \quad i =1,2\,.\label{canonical_transf_2}
\end{eqnarray}
Then, substituting the new variables in $\Hc$ we obtain a Hamiltonian of the form
\begin{eqnarray}
    \tilde{\Hc} - V_0 =  -\frac{\zeta_- }{2} (\tilde{Q}_1^2 +\tilde{P}_1^2) + \frac{\zeta_+}{2}  (\tilde{Q}_2^2 +\tilde{P}_2^2)+\tilde{U}(\tilde{Q}_1,\tilde{Q}_2)\,,
\end{eqnarray}
which has the standard form for a negative energy mode that corresponds to the frequency $\zeta_1:=\zeta_-$.

\bibliographystyle{jpp}
\bibliography{biblio.bib}

\end{document}